\documentclass[journal]{IEEEtran}

\ifCLASSINFOpdf
\else
\fi

\usepackage{graphicx}
\usepackage{threeparttable}
\usepackage{booktabs}
\usepackage{amsmath,amssymb,amsfonts}
\ifCLASSOPTIONcompsoc
\usepackage[caption=false,font=normalsize,labelfont=sf,textfont=sf]{subfig}
\else
\usepackage[caption=false,font=footnotesize]{subfig}
\fi
\usepackage{url}
\usepackage{cite}
\usepackage{amsmath,amssymb,amsfonts,amsthm}
\usepackage[linesnumbered,ruled,vlined]{algorithm2e}

\usepackage[ruled,vlined]{algorithm2e}
\usepackage{algpseudocode}
\usepackage{textcomp}
\usepackage{xcolor}
\usepackage{multirow}

\begin{document}
%
\title{Efficient ML-DSA Public Key Management Method with Identity for PKI and Its Application}
%
%
%

\author{Penghui~Liu,
	Yi~Niu,
	Xiaoxiong~Zhong,
	and Jiahui~Wu,~\IEEEmembership{Member,~IEEE}%
    , Weizhe~Zhang,
    and Kaiping Xue,~\IEEEmembership{Senior Member,~IEEE}%
    , Bin Xiao,~\IEEEmembership{Fellow,~IEEE}%

\thanks{This work is supported in part by the Major Key Project of PCL under Grant PCL2025A13. Penghui Liu (liupenghui1982@163.com and liuph@pcl.ac.cn), Xiaoxiong Zhong (zhongxx@pcl.ac.cn), Jiahui Wu (jiahuiwu2022@163.com), and Weizhe Zhang (weizhe.zhang@pcl.ac.cn) are with the Department of New Networks, Pengcheng Laboratory, Shenzhen, China.
	
Y. Niu is with the Founder of Nanjing Xunshi Data Technology Co., Ltd (niuyi@inssec.cn).

Kaiping Xue is with the School of Cyber Science and Technology, University of Science and Technology of China, Hefei, Anhui 230027, China, and also with Hefei National Laboratory, Hefei, Anhui 230088, China (e-mail: kpxue@ustc.edu.cn).

Bin Xiao is with the Department of Computing, Hong Kong Polytechnic University, Hong Kong (e-mail: csbxiao@comp.polyu.edu.hk).\protect\\
}
\thanks{Manuscript received XXX; revised XXX.}}

\maketitle

\begin{abstract}
With the rapid evolution of the Industrial Internet of Things (IIoT), the boundaries and scale of the Internet are continuously expanding. Consequently, the limitations of traditional certificate-based Public Key Infrastructure (PKI) have become increasingly evident, particularly in scenarios requiring large-scale certificate storage, verification, and frequent transmission. These challenges are expected to be further amplified by the widespread adoption of post-quantum cryptography.
In this paper, we propose a novel identity-based public key management framework for PKI based on post-quantum cryptography, termed \textit{IPK-pq}. This approach implements an identity key generation protocol leveraging NIST ML-DSA and random matrix theory. Building on the concept of the Composite Public Key (CPK), \textit{IPK-pq} addresses the linear collusion problem inherent in CPK through an enhanced identity mapping mechanism. Furthermore, it simplifies the verification of the declared public key's authenticity, effectively reducing the complexity associated with certificate-based key management. We also provide a formal security proof for \textit{IPK-pq}, covering both individual private key components and the composite private key.
To validate our approach, formally, we directly implement and evaluate \textit{IPK-pq} within a typical PKI application scenario: Resource PKI (RPKI). Comparative experimental results demonstrate that an RPKI system based on \textit{IPK-pq} yields significant improvements in efficiency and scalability. These results validate the feasibility and rationality of \textit{IPK-pq}, positioning it as a strong candidate for next-generation RPKI systems capable of securely managing large-scale routing information.

\end{abstract}


%
\IEEEpeerreviewmaketitle

\section{Introduction}\label{section1}
\IEEEPARstart{O}{ver} the past three decades, Public Key Infrastructure (PKI) has been extensively deployed to address a wide range of information security challenges across diverse domains, including e-government, finance, and enterprise operations \cite{Tian}. However, with the rapid emergence of the Industrial Internet of Things (IIoT), the scope of authentication has expanded beyond human entities to encompass a myriad of edge devices, such as sensors. These devices typically operate under severe resource constraints, characterized by narrowband communication, limited computational power, and restricted storage capacity, posing significant challenges to traditional certificate-based management mechanisms \cite{XuS, Turnip}. 
In addition, the continuous expansion of critical internet infrastructure has introduced scalability bottlenecks to large-scale certificate management. These limitations are expected to be significantly exacerbated by the anticipated widespread adoption of post-quantum cryptography (PQC). While a direct ``drop-in'' replacement of traditional algorithms with NIST-standardized post-quantum primitives (e.g., ML-DSA) appears to be a straightforward migration path, it introduces prohibitive overheads for resource-constrained environments. Specifically, the inherent key and signature sizes of post-quantum algorithms are substantially larger than their classical counterparts. For example, a standard X.$509$ format certificate based on ML-DSA-$87$ requires approximately $10$ KB of storage space, while a typical RSA-$2048$ certificate only has about $1$ KB. This nearly tenfold increase, often referred to as an ``order-of-magnitude'' expansion, becomes catastrophic in scenarios relying on deep certificate chains. Consequently, the cumulative overhead of transmitting and verifying multiple ``heavyweight'' post-quantum certificates inevitably leads to unacceptable bandwidth consumption and processing latency. Moreover, directly integrating post-quantum primitives into existing protocols would introduce significant complexity and verification challenges, as demonstrated in recent formal analyses of post-quantum key agreement protocols \cite{Bhargavan}. Chhetri et al. \cite{ChhetriG} pointed out the practical deployment challenges after NIST standardization (FIPS $203/204/205$), especially the "hard to implement�problem in the Internet of Things (IoT) and critical infrastructure. Therefore, exploring a novel public key management mechanism that eliminates the dependency on heavy certificate chains is a critical research imperative for the post-quantum era.

This scalability challenge is particularly acute in the Resource Public Key Infrastructure (RPKI), where certificate chains are inherently deep due to its hierarchical governance model. As illustrated in Figure \ref{figure 5}, the RPKI system operates within a hierarchical resource management framework. The Internet Assigned Numbers Authority (IANA) is positioned at the apex, followed by five Regional Internet Registries (RIRs), such as the Asia Pacific Network Information Center (APNIC), tasked with managing Internet Number Resources (INRs), including IP addresses and Autonomous System (AS) numbers within their respective regions. The third layer comprises National Internet Registries (NIRs), Local Internet Registries (LIRs), and independent IP address holders, while in some regions, this layer may consist directly of Internet Service Providers (ISPs) \cite{6RFC6480}. Within this framework, INR holders delegate portions of their IP address blocks to subordinate entities using X.$509$-formatted Resource Certificates (RCs). These certificates utilize extensions to transmit IP authorization information, ensuring the consistency of resource allocation and the security of the Internet routing infrastructure \cite{APNICRFC7935}. The primary objective of RPKI is to bind INRs to a trust anchor, enabling relying parties (RPs) to verify the authenticity of Border Gateway Protocol (BGP) routes, thereby mitigating routing prefix hijacking attacks \cite{HFRPKI}.

According to the stipulation in RFC$7935$ \cite{APNICRFC7935},
the global RPKI specifies the selection of cryptographic algorithms, limiting the signature algorithm to RSA2048 and the HASH algorithm to SHA$256$.
However, RPKI exhibits the following limitations: ($1$) Given the massive scale of global IP address resources, issuing RC (Resource certificate) and other related objects e.g., ROA (Route Origin Authorization), CRL (Certificate Revocation List), and MFT (Manifest) requires substantial computation \cite{IPAS}. ($2$) Moreover, RPKI relies on recipients (relying parties or other third parties) to pre-download multiple RCs and construct complete certificate chains during large-scale routing updates, further increasing the verification overheads \cite{RPKIRFC}. ($3$) As a global critical infrastructure maintaining numerous signed objects, RPKI imposes significant computational burdens, especially for signature verification. This is particularly challenging for edge network devices with limited resources, where large-scale verification consumes substantial CPU cycles, even with hardware acceleration \cite{RakshaCarl}. ($4$) Additionally, the use of Rsync or RRDP protocols for periodic synchronization of certificate chains and signed objects inevitablely increase the workload of the control plane \cite{Bruijnzeels}. Therefore, enhancing RPKI performance and exploring alternative standardized cryptographic solutions, particularly for the post-quantum era, is a key research direction for internet infrastructure.

\textbf{Related Work}.
To reduce the complexity and overhead of certificate-based public key management in PKI, various identity-based and CPK-based cryptographic schemes, such as SM9 \cite{HJI}, PKIot \cite{MarinoF}, IBE-Lite \cite{TanCC, GeC, WangH}, IMS \cite{ShenJ}, PKI4IoT \cite{Hoglund}, and CPK \cite{HaberS, NanX2, IPK}, have been proposed to enhance security in IoT and cloud environments. Gavriilidis et al.\cite{Gavriilidis} also summarized and proved that the IBC scheme has significant advantages over traditional PKI schemes in decryption time and communication overhead (reducing communication traffic by about 18\%). However, these identity-based and CPK schemes primarily rely on traditional RSA/ECC cryptosystems, rendering them unsuitable for post-quantum PKI systems. In the context of RPKI, limited research has focused on reducing cryptographic overhead, with most efforts instead targeting the regulation of RIRs, decentralization to mitigate their authority, performance measurement, monitoring, and resource security audits \cite{HZou, Roachain, RPKISurvey}. Recent research \cite{KangM} provides an in-depth analysis of RPKI and IRR data, revealing operational inconsistencies between RPKI and legacy IRR records, and the existing issues in RPKI deployment include not only large certificate storage space but also chaotic data management, highlighting the need for a more robust and automated management framework. Some works in TIFS have explored alternatives to traditional PKI for post-quantum IoT environments. For instance, Xu et al. \cite{XuS} proposed a post-quantum certificateless signcryption scheme to eliminate certificate management overheads in the Internet of Medical Things (IoMT). While their approach primarily addresses the key escrow problem associated with identity-based cryptography, it is not suitable for Public Key Infrastructure (PKI) systems, our work specifically targets the PKI infrastructure by leveraging matrix-based identity mapping to achieve stricter verification efficiency. Kyung-Ah Shim et al. \cite{KyungAhShim} proposed a post-quantum identity-based signature authentication framework for IoT smart devices. The core technology focuses on directly utilizing existing post-quantum signature primitives, such as lattice-based, hash-based, and code-based schemes, to construct a lightweight, certificate-free authentication framework. This framework is not designed to adapt to existing Public Key Infrastructures (PKI) but provides an innovative approach for achieving post-quantum secure authentication in resource-constrained IoT environments, effectively bypassing the complexities of PKI. It is particularly well-suited for deployment in next-generation IoT security systems and is not applicable to the PKI systems. Turnip et al. \cite{Turnip} pointed out that integrating PQC directly into protocols such as TLS and SSH would result in a $600\%$ increase in handshake latency and introduce significant memory overhead. To address the current 5G-AKA protocol's lack of perfect forward secrecy and post-quantum (PQ) security, Braeken et al. \cite{Braeken} proposed a solution based on post-quantum techniques, which directly employs PQ encryption primitives for confidentiality and privacy protection while retaining classical public key cryptography for authentication. However, this solution is not suitable for PKI. Some studies have proposed modifying existing PKI systems by integrating post-quantum algorithms recommended by NIST, such as ML-DSA (e.g., Crystals-Dilithium), which can be applied to RPKI \cite{Raavi, DOnghia} and are referred to here as the ``standard scheme".

\textbf{Contributions}.
In this work, we address the critical performance and efficiency bottlenecks in RPKI by proposing \textit{IPK-pq}, a novel framework that integrates CPK technology with random matrix theory. This approach minimizes modifications to the existing BGP and RPKI architecture while significantly enhancing the operational efficiency of ML-DSA-based systems. To evaluate its feasibility, we constructed a comprehensive test environment to benchmark \textit{IPK-pq} against the standard scheme, assessing its adaptability within RPKI ecosystems and conducting rigorous stress testing on critical operations, such as ROA signing and verification. Based on these results, we further analyze its potential for deployment in real-world production environments. In summary, we make the following main contributions:

\begin{itemize}
	\item We propose \textit{IPK-pq}, the first identity-based and HSM (Hardware security module)-supported post-quantum PKI scheme. By leveraging matrix-based identity mapping, it eliminates the dependency on deep certificate chains, streamlining public key management and effectively resolving the long-standing key escrow challenge inherent in traditional IBC.
	
	\item We provide formal proofs demonstrating that \textit{IPK-pq} is resilient against linear collusion and public key substitution attacks, while ensuring robust collision resistance for private keys.
	
	\item We experimentally ported \textit{IPK-pq} to the NLnetLabs' open-source Krill and Routinator software, implementing a fully functional prototype for the issuance and verification of RCs and signed objects (e.g., ROAs).
	
	\item We conducted a comprehensive comparative analysis, which demonstrates that \textit{IPK-pq} achieves superior amortized communication efficiency. Specifically, our results show that \textit{IPK-pq} reduces the communication complexity of verification from $\mathcal{O}(n)$ (linear growth with CA depth) to $\mathcal{O}(1)$ (constant), making it significantly more efficient and scalable than the standard scheme.
	
	\item We adapted the \textit{IPK-pq}-based RPKI to the NXP Freescale® C$293$ PCIe Card. The performance analysis confirms that this hardware-accelerated solution is fully suitable for large-scale, high-throughput RPKI scenarios.
\end{itemize}

\textbf{Outline}. We provide preliminaries in Section~\ref{section2}. We elaborate on the system and threat model, and describe our proposed RPKI system in Section~\ref{section3}. Section~\ref{section4} gives the theoretical security analysis of \textit{IPK-pq}. Then we discuss the performance and analysis of our RPKI system in Section~\ref{section5}, and we conclude in Section~\ref{section6}.

\section{Preliminaries}\label{section2}
\subsection{Resource Public Key Infrastructure}
RPKI primarily consists of four components:  CA, Relying party, Repository, and BGP router. Among them, CA and Repository can be deployed in one management entity \cite{MADRPKI, IETFRFC6916}. These components work together to manage the INRs of the Internet by issuing, transmitting, storing, and verifying various RCs and digital objects, including ROAs and MFTs \cite{BBN}. 
These components are briefly described as follows:
\begin{itemize}
	\item \textbf{CA} issues the RCs and related digital objects (e.g., CRLs) for subordinate CA to indicate the allocation information of INR or directly issues ROAs to authorize ISP to announce ownership of the specific IP address prefixes, and publishes the RCs and digital objects into database of the Repository managed by itself or parent CA.
	\item \textbf{Relying Party (RP)} retrieves the RCs and digital objects from the Repository via Rsync or RRDP protocols and processes them to establish the verified and authentic authorization relationship between IP address blocks and AS numbers, which can be utilized to verify the BGP routing information for the BGP router.
	\item \textbf{Repository (Repo.)} is a database that stores the issued RCs and digital objects, which can be downloaded by any third party, including the relying party. Moreover, Repo. is also responsible for maintaining the historical versions and downloading information of RCs and digital objects.
	\item \textbf{BGP Router} is designed to collect the validated ROA payloads from one or more Relying-party sources to acquire the real authorization relationships via RTR protocols, to check the validity of route information after obtaining the routing advertisement.
\end{itemize}

\begin{figure}[!t]
	\centering
	\includegraphics[width=\linewidth]{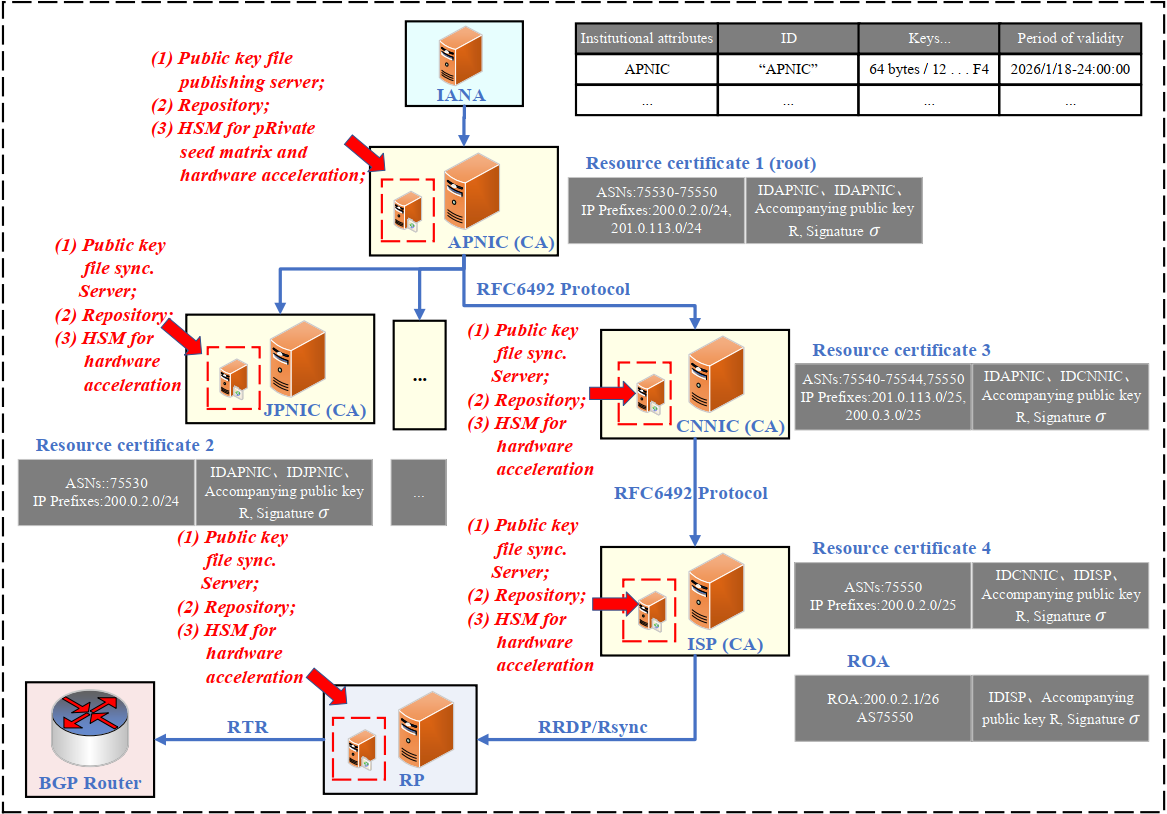}
	\caption{RPKI resource management technology based on IPK-pq}
	\label{figure 5}
\end{figure}

\subsection{Identity-based Cryptography Theory and Technology}
Identity-based cryptography \cite{IBC} leverages a user's public identity (e.g., unique name, email, or phone number) as the public key, while the private key is generated by a Key Generation Center (KGC) using a master key and the user's identity, thereby eliminating the need for certificate management in PKI. \textit{IPK-pq} is an identity-based version of ML-DSA \cite{MLDSA}, which is a novel key management methodology that builds upon the concept of CPK and resolves the linear collusion issue related to CPK, as well as the authenticity and verifiability problem of the declared public key by enhancing the identity mapping approach.

\section{Certificate-chain-free RPKI: IPK-pq}\label{section3}
\subsection{Model and Basic Algorithms}
The main notations used in this paper are summarized in Table \ref{Notations}, other notations for ML-DSA can be found in \cite{MLDSA}. \textit{IPK-pq} shares the same management architecture as the current RPKI. Its threat model includes external adversaries but excludes participating entities, such as RIRs/NIRs. While the key pairs refreshed or rotated by both mechanisms are similar, they differ in the way they bind the public key: the current PKI utilizes certificates, whereas \textit{IPK-pq} employs identities. As an instance of the CPK methodology, \textit{IPK-pq} leverages a compact seed matrix and identity mapping, coupled with mathematical algorithms, to establish identity-key bindings. This approach replaces certificate-based public key management with a more efficient identity-based framework. Consequently, \textit{IPK-pq} introduces a streamlined, certificate-chain-free system that simplifies the management of large public key sets for users and devices. Its core functionality is defined by the following three fundamental algorithms:

\begin{table}[!tbp]
	\centering
	\caption{List of Notations }\label{Notations}
	\begin{threeparttable}
		\begin{tabular}{p{0.1\linewidth}p{0.8\linewidth}}
			\toprule
			Notations & Description \\
			\midrule
			$m$ & the row count of seed matrix $\mathbf{M}_{\rho'}^{\text{piv}}$ and $\mathbf{M}_{\rho}^{\text{pub}}$. \\
            \hline
			$h$ & the column count of seed matrix $\mathbf{M}_{\rho'}^{\text{piv}}$ and $\mathbf{M}_{\rho}^{\text{pub}}$. \\
            \hline
			$ID_{\mathsf{CA}}$ & a public identity for CA, owned by CA. \\
			\hline
			$R_{\mathsf{CA}}$ & a 32-byte accompanying public key for CA, owned by CA.  \\
			\hline
			$isk_{\mathsf{CA}}$ & the ML-DSA private key for CA, owned by CA.  \\
			\hline
			$IPK_{\mathsf{CA}}$ & the ML-DSA public key for CA, owned by CA.  \\
            \hline
			$K_{\mathsf{CA}}$ & a 32-byte ML-DSA private random seed for use in signing, owned by CA. \\
			\hline
			$\rho_{\mathsf{CA}}$ & a 32-byte ML-DSA public random seed, owned by CA. \\
			\hline
			$\ensuremath{\rho}'_{\mathsf{CA}}$ & a 64-byte ML-DSA private random seed, owned by CA. \\
            \hline
			$K_{\mathsf{KC}}$ & a 32-byte ML-DSA private random seed for use in signing, owned by Key Center. \\
			\hline
			$\ensuremath{\rho}_{\mathsf{KC}}$ & a 32-byte ML-DSA public random seed, owned by Key Center. \\
			\hline
			$\ensuremath{\rho}'_{\mathsf{KC}}$ & a 64-byte ML-DSA private random seed, owned by Key Center. \\			
			\hline
            $\mathbf{M}_{\rho'}^{\text{piv}}$ & Private $m\times h$ random seed matrix for mapping, containing $m\times h$ random seed $ \rho'$, stored in the Key Center's HSM.  \\
            \hline
            $\mathbf{M}_{\rho}^{\text{pub}}$ & Public $m\times h$ random seed matrix for mapping, containing $m\times h$ random seed $ \rho$, stored in the public key file. \\
			\hline
           	$\ensuremath{\rho}'_{\mathsf{r(KC)}}$ & a 64-byte private random seed for CA registration, owned by Key Center; different CAs employ different random values. \\
			\hline
            $\ensuremath{\rho}'_{\mathsf{r(CA)}}$ & a 64-byte private random seed for CA registration, owned by CA. \\
			\hline
            $r_{\mathsf{CA}}$ & a 32-byte random number for CA registration, owned by CA. \\
			\hline
			 $A_{\mathsf{CA}}$ & a polynomial matrix is sampled from $R^{k\times l}_q$ based on $\rho_{\mathsf{CA}}$, owned by CA.\\
			\hline
			$File_{\mathsf{PK}}$ & a public key file containing $\mathbf{M}_{\rho}^{\text{pub}}$ and the public key record $PK_{\mathsf{KC}}[ID_{\mathsf{CA}}]$ for all CAs, maintained by Key Center.  \\			
            \bottomrule			
		\end{tabular}
	\end{threeparttable}
\end{table}

\begin{figure}[h]
	\centering
	\includegraphics[width=\linewidth]{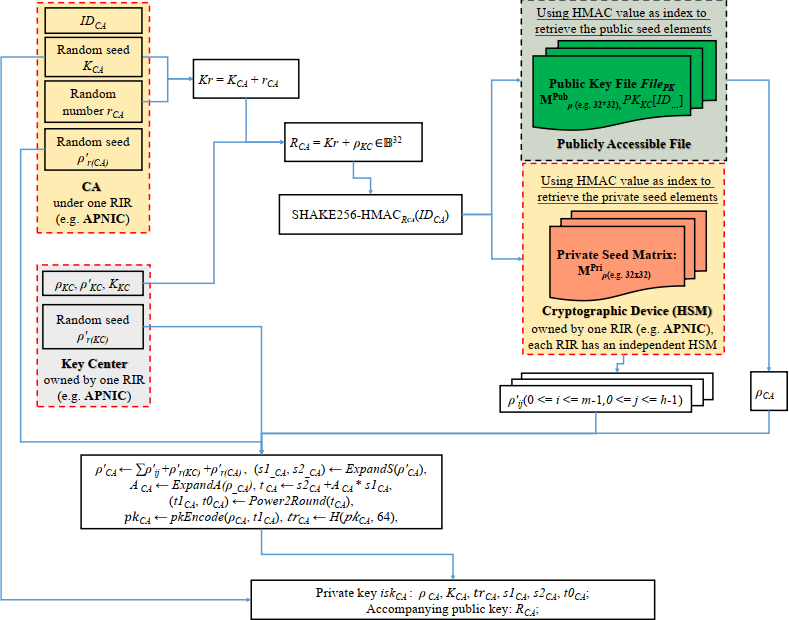}
	\caption{IPK-pq CA private key generation process}
	\label{figure 2a}
\end{figure}

\begin{figure}[h]
	\centering
	\includegraphics[width=\linewidth]{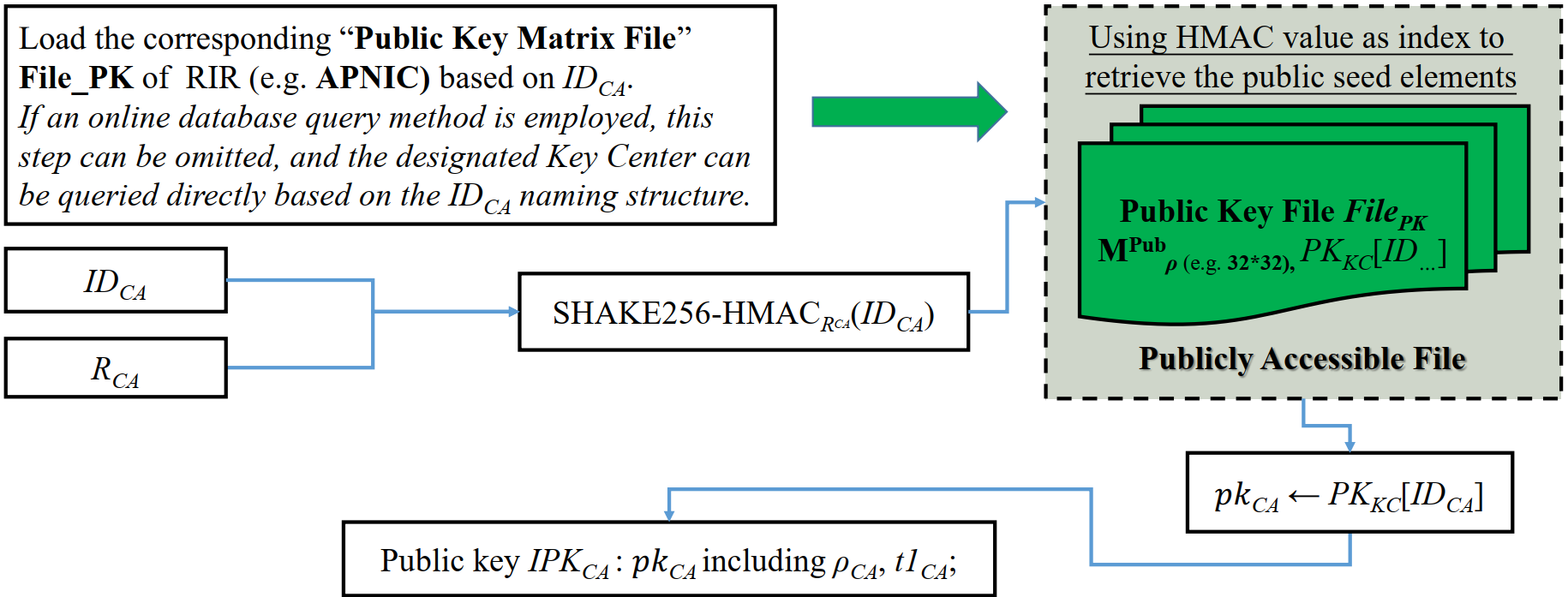}
	\caption{IPK-pq CA self-service public key generation}
	\label{figure 2b}
\end{figure}

\subsubsection{\textbf{System Private Key Seed Matrix Generation}}
During system deployment phase, HSM generates $m\times h$ (e.g., $32\times32$, $m = 32, h = 32$) private random seeds of ML-DSA using Pseudo Random Number Generator (PRNG), stores these $mh$ $64$-byte private random seeds into secure and confidential area of cryptographic device in order, forming the private random seed matrix $\mathbf{M}_{\rho'}^{\text{piv}}$, then similarly generates the ML-DSA public random seed matrix $\mathbf{M}_{\rho}^{\text{pub}}$, and outputs $\mathbf{M}_{\rho}^{\text{pub}}$ into public key file $File_{\mathsf{PK}}$.

\begin{algorithm}[!t]
	\caption{Retrieving the ML-DSA public random seed $\rho_{\mathsf{CA}}$ from public random seed matrix $\mathbf{M}_{\rho}^{\text{pub}}$ } \label{PARTPK_ACCESS}
	\KwIn{The identity $ID_{\mathsf{CA}}$ of CA and its accompanying public key $R_{\mathsf{CA}}$; the public random seed matrix $\mathbf{M}_{\rho}^{\text{pub}}$.}
	\KwOut{The ML-DSA public random seed $\rho_{\mathsf{CA}}$.}
	Initializes $\ensuremath{\rho}_{\mathsf{CA}} \leftarrow 0$\;
 	Generates a $256$ bits $hmac \leftarrow \textsf{SHAKE256-HMAC}_{R_{\mathsf{CA}}}(ID_{\mathsf{CA}})$ \tcc*{Calculate the random index for retrieving}
    Evenly divides $hmac$ \textbf{into} $h$ sequential segments $hmac[i](i = 0, 1, \cdots, h-1)$\;
	\For {$col$ $\in$ \{$0, 1, 2, \cdots, h-1$\}} {
		$row \leftarrow \textsf{BitsToInteger}(hmac[col])\ mod\ m$\;
		$\ensuremath{\rho}_{temp} \leftarrow \mathbf{M}_{\rho}^{\text{pub}}[row][col]$ \tcc*{Retrieves the public random seed.}
		$\ensuremath{\rho}_{\mathsf{CA}} \leftarrow \ensuremath{\rho}_{\mathsf{CA}} + \ensuremath{\rho}_{temp}$\;
	}
	\Return $\rho_{\mathsf{CA}}$\;
\end{algorithm}

\subsubsection{\textbf{Collaborative Private Key Generation}}

\begin{algorithm}[!t]
	\caption{\textit{IPK-pq} Private Key Generation Process}\label{IPK_PRIV}
	\KwIn{The identity $ID_{\mathsf{CA}}$ of CA, the ML-DSA private random seed matrix $\mathbf{M}_{\rho'}^{\text{piv}}$, the public key file $File_{\mathsf{PK}}$.}
	\KwOut{The ML-DSA private key $isk_{\mathsf{CA}}$ for CA, the accompanying	public key $R_{\mathsf{CA}}$.}
	CA generates a private ML-DSA random seed $K_{\mathsf{CA}}$ $\leftarrow$ $\textsf{PRNG}(\cdot)$, $\ensuremath{\rho}'_{\mathsf{r(CA)}}$ $\leftarrow$ $\textsf{PRNG}(\cdot)$, and a random number $r_{\mathsf{CA}}$ $\leftarrow$ $\textsf{PRNG}(\cdot)$, then computes $Kr \leftarrow $ $K_{\mathsf{CA}} $ $+$ $r_{\mathsf{CA}}$, and sends $Kr$ to Key Center\;
	Key Center allocates and initializes a public key element $t1_{\mathsf{KC}}[ID_{\mathsf{CA}}]$ $\leftarrow$ $ 0$, then generates a private random seed $\ensuremath{\rho}'_{\mathsf{r(KC)}}$ $\leftarrow$ $\textsf{PRNG}(\cdot)$,
	and computes an accompanying public key $R_{\mathsf{CA}} \leftarrow Kr+\ensuremath{\rho}_{\mathsf{KC}}$ for CA with its public ML-DSA random seed $\ensuremath{\rho}_{\mathsf{KC}}$\;
	Key Center (HSM) initializes $\ensuremath{\rho}'_{\mathsf{CA}} \leftarrow 0$, $\ensuremath{\rho}_{\mathsf{CA}} \leftarrow 0$\;		
    Key Center loads the ML-DSA public random seed matrix $\mathbf{M}_{\rho}^{\text{pub}}$ from $File_{\mathsf{PK}}$, runs \textbf{Algorithm \ref{PARTPK_ACCESS}} to retrieve the ML-DSA public random seed $\rho_{\mathsf{CA}}$ for CA with parameters $ID_{\mathsf{CA}}$, $R_{\mathsf{CA}}$, and $\mathbf{M}_{\rho}^{\text{pub}}$ \tcc*{Generates the ML-DSA public random seed for CA.}
    Key Center (HSM) generates a $256$ bits $hmac \leftarrow \textsf{SHAKE256-HMAC}_{R_{\mathsf{CA}}}(ID_{\mathsf{CA}})$ \tcc*{Generates the ML-DSA private key for CA.}
    Key Center (HSM) evenly divides $hmac$ into $h$ sequential segments $hmac[i](i = 0,1,...,h-1)$\;
	\For {$col$ $\in$ \{$0, 1, 2, \cdots, h-1$\}} {
         $row \leftarrow \textsf{BitsToInteger}(hmac[col])\ mod\ m$\;
         $\ensuremath{\rho}'_{temp} \leftarrow \mathbf{M}_{\rho'}^{\text{piv}}[row][col]$ \tcc*{Fetches the private random seed located at Row row, Column col.}
	 	 $\ensuremath{\rho}'_{\mathsf{CA}} \leftarrow \ensuremath{\rho}'_{\mathsf{CA}} + \ensuremath{\rho}'_{temp}$\;
	}
	Key Center computes:
	$\ensuremath{\rho}'_{\mathsf{CA}} \leftarrow \ensuremath{\rho}'_{\mathsf{CA}}+\ensuremath{\rho}'_{\mathsf{r(KC)}}$;
	then sends  $R_{\mathsf{CA}}$, $\ensuremath{\rho}'_{\mathsf{CA}}$ and $\rho_{\mathsf{CA}}$ to CA\;
	CA computes:	 	
	$\ensuremath{\rho}'_{\mathsf{CA}} \leftarrow \ensuremath{\rho}'_{\mathsf{CA}}+\ensuremath{\rho}'_{\mathsf{r(CA)}}$;
	$A_{\mathsf{CA}} \leftarrow \mathbf{ExpandA}(\ensuremath{\rho}_{\mathsf{CA}})$;
	$(s1_{\mathsf{CA}}, s2_{\mathsf{CA}}) \leftarrow \mathbf{ExpandS}(\ensuremath{\rho}'_{\mathsf{CA}})$;
	$t_{\mathsf{CA}} \leftarrow s2_{\mathsf{CA}} +A_{\mathsf{CA}}\times s1_{\mathsf{CA}}$;
	$(t1_{\mathsf{CA}}, t0_{\mathsf{CA}}) \leftarrow \mathbf{Power2Round}(t_{\mathsf{CA}})$;
    then sends $t1_{\mathsf{CA}}$ to Key Center\;
	Key Center computes $pK_{\mathsf{CA}} \leftarrow \mathbf{pkEncode}(\ensuremath{\rho}_{\mathsf{CA}}, t1_{\mathsf{CA}})$, then writes $PK_{\mathsf{KC}}[ID_{\mathsf{CA}}] \leftarrow pK_{\mathsf{CA}}$ into public key file $File_{\mathsf{PK}}$ in an append mode\;
	CA computes the private key:
	$pK_{\mathsf{CA}} \leftarrow \mathbf{pkEncode}(\ensuremath{\rho}_{\mathsf{CA}}, t1_{\mathsf{CA}})$;
	$tR_{\mathsf{CA}} \leftarrow H(pK_{\mathsf{CA}}, 64)$;
	$isk_{\mathsf{CA}}$ $\leftarrow$ $\mathbf{skEncode}(\ensuremath{\rho}_{\mathsf{CA}},K_{\mathsf{CA}},$$ tR_{\mathsf{CA}}, s1_{\mathsf{CA}}, s2_{\mathsf{CA}}, t0_{\mathsf{CA}})$\;
	\Return $isk_{\mathsf{CA}}$, $R_{\mathsf{CA}}$\;
\end{algorithm}

The generation of a CA's private key is collaboratively performed by the CA and the RIR's Key Center, which operates an HSM. This ensures that only the CA has access to its complete private key, while it remains inaccessible to the Key Center. The decentralized private key generation process addresses the limitation of relying solely on the Key Center for private key generation, ensuring compliance with electronic signature laws. The private key generation process is outlined in Algorithm \ref{IPK_PRIV} and Figure \ref{figure 2a}. During the private key generation for CA $[ID_{\mathsf{CA}}]$, the Key Center also computes the CA's public key and establishes its index $[ID_{\mathsf{CA}}]$ for efficient retrieval, as described in Algorithm \ref{IPK_PRIV} \footnote{The functions referenced in Algorithms \ref{PARTPK_ACCESS} and \ref{IPK_PRIV} are detailed in NIST FIPS 204 \cite{MLDSA}, where $\textsf{SHAKE256-HMAC}$ represents an HMAC algorithm based on $\textsf{SHAKE256}$.}. Finally, the Key Center appends the CA's public key to the public key file $File_{\mathsf{PK}}$, compresses it, and stores it on a powerful, publicly accessible publishing server \footnote{An online database query service of public key can also be provided to eliminate the need for file downloads, depending on the application scenario.}.
Upon completion of this stage, the private key obtained by the CA is $isk_{\mathsf{CA}}$, and the explicit public key is \{$ID_{\mathsf{CA}}$, $R_{\mathsf{CA}}$\}.


\begin{algorithm}[!t]
	\caption{\textit{IPK-pq} Public Key Generation Process} \label{PUB_ACCESS}
	\KwIn{The identity $ID_{\mathsf{CA}}$ of CA, its accompanying public key $R_{\mathsf{CA}}$, the public key file $File_{\mathsf{PK}}$.}
	\KwOut{The ML-DSA public key $IPK_{\mathsf{CA}}$ for CA.}
	Initializes $\ensuremath{\rho}_{\mathsf{CA}} \leftarrow 0$\;
	Loads the ML-DSA public random seed matrix $\mathbf{M}_{\rho}^{\text{pub}}$ from $File_{\mathsf{PK}}$\;
	Runs \textbf{Algorithm \ref{PARTPK_ACCESS}} to retrieve the ML-DSA public random seed $\rho_{\mathsf{CA}}$ for CA with parameters $ID_{\mathsf{CA}}$, $R_{\mathsf{CA}}$, and $\mathbf{M}_{\rho}^{\text{pub}}$
	\tcc*{Generate the ML-DSA public random seed for CA.}
    Retrieves $PK_{\mathsf{KC}}[ID_{\mathsf{CA}}]$ with $ID_{\mathsf{CA}}$ from the public key file $File_{\mathsf{PK}}$, $pK_{\mathsf{CA}} \leftarrow PK_{\mathsf{KC}}[ID_{\mathsf{CA}}]$\;
	Decodes $\ensuremath{\rho}''_{\mathsf{CA}}$ from $pK_{\mathsf{CA}}$ in $PK_{\mathsf{KC}}[ID_{\mathsf{CA}}]$\;
	Checks whether $\ensuremath{\rho}''_{\mathsf{CA}}$ is equal to $\rho_{\mathsf{CA}}$ retrieved above\;
    \Return $IPK_{\mathsf{CA}} \leftarrow pK_{\mathsf{CA}}$ if yes; otherwise ``$\perp$".\
\end{algorithm}

\subsubsection{\textbf{Self-Service Public Key Generation}}
The calculation of a CA's ML-DSA public key, based on its identity $ID_{\mathsf{CA}}$, accompanying public key $R_{\mathsf{CA}}$, and the public key file $File_{\mathsf{PK}}$, can be performed by any participant (CA or RP) in the RPKI system as needed. The resulting public key $IPK_{\mathsf{CA}}$, paired with the private key $isk_{\mathsf{CA}}$, forms an ML-DSA key pair. The coordinate selection in the binding generation matrix between the CA's identity and the accompanying public key addresses the substitution attack problem associated with the declared public key. The process for calculating the CA's ML-DSA public key is shown in Figure \ref{figure 2b} and detailed in Algorithm \ref{PUB_ACCESS}.

It is clear that \textit{IPK-pq} establishes a direct binding between a CA's identity and its public key by mapping the identity to the key seed matrix, with $R_{\mathsf{CA}}$ serving as proof. This approach simplifies the management of large numbers of public keys for users or devices by equating the identity with the public key. \textit{IPK-pq} generates key pairs structurally identical to those of the standard ML-DSA scheme, with the key difference lying in their management mechanisms: \textit{IPK-pq} uses the identity directly as the public key, while the standard scheme relies on certificates for key management. Since both systems produce equivalent ML-DSA key pairs, they ensure interoperability and compatibility in cryptographic operations.

\subsection{System Setup Stage}
The system framework of \textit{IPK-pq} is illustrated in Figure \ref{figure 5}. Generally, IANA delegates INR resources to five regional RIRs, each maintaining a trust anchor. In our work, the RIR serves as the Key Center, acting as the authority for regional resource management. It oversees resource allocation and initializes the private key seed matrix in the HSM following \textit{``System Private Key Seed Matrix Generation"}. There are five RIRs that receive corresponding resource allocations from IANA, with Figure \ref{figure 5} focusing solely on APNIC in the Asia Pacific region. This section uses APNIC as an example to illustrate our work. As the root CA, APNIC oversees key management for all regional CAs in the Asia Pacific region. The CA naming convention is defined as follows.
\begin{itemize}
	\item The APNIC CA serves as the root CA in the Asia Pacific region, could be named $ID_{\mathsf{APNIC}}$ , such as ``$APNIC$".
	\item The naming of each subordinate CA of RIR follows the same pattern. As shown in Figure \ref{figure 5}, CNNIC, located under APNIC, is China Internet Network Information Center, currently responsible for managing the INRs in the Chinese Mainland, and its CA could be named $ID_{\mathsf{APNIC}}||ID_{\mathsf{CNNIC}}$, i.e. ``$APNIC||CNNIC$".
	\item Each ISP CA under CNNIC can follow a consistent naming pattern. For example, if a Chinese ISP applies for INRs from CNNIC and establishes a CA to manage them, its name could be: $ID_{\mathsf{APNIC}}||ID_{\mathsf{CNNIC}}||ID_{\mathsf{ISP}}$.
\end{itemize}
Additionally, if an ISP has sub-CAs, they can be named using a similar hierarchical naming convention. Notably, APNIC employs a dedicated HSM to store a secure, randomly generated private seed matrix, which is inaccessible and updatable as needed to align with management policies, ensuring strong security and compliance with governance standards.


\subsection{Registration Stage}
Before requesting INR resources from their parent CA, all regional CAs in the Asia Pacific region, including APNIC, must submit a registration request to the APNIC CA in accordance with established procedures and governance policies. This process can be conducted either online or offline, which is not constrained by this work. Upon receiving the registration request, APNIC and the CA jointly follow the procedure shown in Figure \ref{figure 2a} \textit{``\textit{IPK-pq} CA private key generation process"}, to generate the corresponding private key and create a registration record, which includes Institutional attributes, Identity $ID$, Random number $\ensuremath{\rho}'_{\mathsf{r(KC)}}$, Accompanying public key $R_{\mathsf{CA}}$, and Period of validity. Then APNIC returns the registration record and the URL of public key file $File_{\mathsf{PK}}$ to the CA. Finally, APNIC publishes the registration information (excluding Random number $\ensuremath{\rho}'_{\mathsf{r(KC)}}$) and public key file to the entire RPKI system, allowing all entities in the system, including CAs and RPs, to synchronize the public key file via the synchronization server\footnote{For devices with limited storage, the system can offer an online database query service of public key, eliminating the need for file downloads.}. After its validity period expires, a CA must renew its registration, update its private key, and synchronize the public key file if needed. Upon completing registration, each CA is issued an accompanying public key $R_{\mathsf{CA}}$. Table \ref{The record of Registration} shows the example records of Registration.

\begin{table*}
	\caption{The Example Record of Registration}
	\label{The record of Registration}
	\centering
	\begin{tabular}{{p{1cm}|p{3cm}|p{4cm}|p{2cm}|p{2cm}|p{3cm}}}
		\toprule
		No.&Institutional\ \ \ \  \ Attributes&$ID_{\mathsf{CA}}$& $\ensuremath{\rho}'_{\mathsf{r(KC)}}$ &$R_{\mathsf{CA}}$ &Period of Validity\\
		\midrule
		1 & APNIC & ``$APNIC$" & $64$B \// $D2 …A3$ & $32$B \// $01 …BD$ & 2026/6/18-24:00:00, 2027/6/18-23:59:59\\
		2 & CNNIC & ``$APNIC||CNNIC$" & $64$B \// $E3 …FB$ & $32$B \// $E2 â€?F$ & 2026/7/18-24:00:00, 2027/7/18-23:59:59\\
		3 & China XXXcom ISP & ``$APNIC||CNNIC||CNXXX$" & $64$B \// $A5 …ED$ & $32$B \// $B5 …D5$ & 2026/8/18-24:00:00, 2027/8/18-23:59:59\\
		4 & ... & ... & ... &  ... & ...\\
		\bottomrule
	\end{tabular}
\end{table*}

\subsection{Resource Allocation and ROA Issuance Stage}
If a regional non-root CA (referred to as CA$1$) is required to obtain INRs from its parent CA (referred to as CA$2$), CA$2$ firstly generates a X.$509$-formatted RC framework and initializes the RC, which encompasses at least the following fields: $[CA2\ name] + [CA1\ name] + [AS/IP\ segment] + [Signature]$.
And then, CA$2$ signs the RC and delivers it to CA$1$. Based on the signer's name specified in the RC, which here corresponds to $[CA2\ name]$, CA$1$ firstly self-calculates the CA$2$'s public key using the procedure depicted in Figure \ref{figure 2b} \textit{``\textit{IPK-pq} CA self-service public key generation"}. Then, ML-DSA is utilized to verify the signature $\sigma$ in $[Signature]$ of the RC. If the verification result is valid, CA$1$ accepts the allocated RC; otherwise, CA$1$ must reapply for INRs as required.

\begin{figure}[h]
	\centering
	\includegraphics[width=\linewidth]{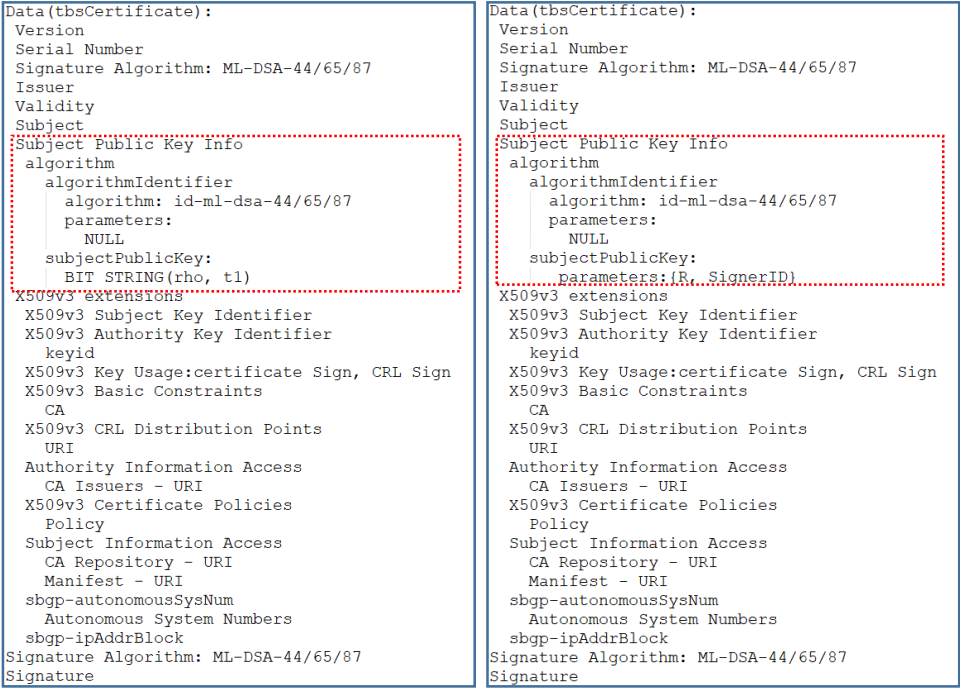}
	\caption{Differences in RC data fields between standard ML-DSA and IPK-pq}
	\label{figure 24}
\end{figure}

Noting that, $[SubjectPublicKeyInfo]$ in the RC is generated using \textit{IPK-pq}, exhibits a slight variation with that of ML-DSA in the standard scheme, it incorporates the signer's identity $ID_{\mathsf{CA}}$ and accompanying public key $R_{\mathsf{CA}}$, which is also shown in Figure \ref{figure 24}\footnote{RC in \textit{IPK-pq} can still associate with its parent CA's specific RC through X.$509$ extensions, indicating INR allocation and inclusion relationships. Issuer is usually the SignerID, which is defined separately to increase the flexibility.}. Hence, CA$1$ eliminates the demand for downloading certificate chain by directly self-calculating the signer's public key to complete the signature verification with public key file $File_{\mathsf{PK}}$, thereby significantly enhancing the efficiency of the verification process.

When the lowest-level CA (referred to as CA$3$) requires to publish ROA records, the process is similar to the above RC request, but with the following ROA fields: $[CA3\ name] + [AS/IP\ segment] + [Signature]$.

\subsection{ROA Verification Stage}
If a third party, including Relying party, intends to verify the signature upon receiving an ROA object, it can typically self-calculate the CA$3$'s public key based on issuer's identity using the algorithm depicted in Figure \ref{figure 2b} \textit{``\textit{IPK-pq} CA self-service public key generation"}. Subsequently, the validity of $[Signature]$ in this ROA object can be directly verified using ML-DSA. This approach eliminates the need for third-party to download the certificate chain containing all RCs from APNIC to the CA entity issuing this ROA and EE certificate, thereby significantly reducing computational and communication overheads.

\section{Theoretical Security Analysis of IPK-pq}\label{section4}
As a public key management approach based on ML-DSA, the theoretically proven security of \textit{IPK-pq} still relies on the problems of Module Learning With Errors (MLWE) and Module Short Integer Solution (MSIS)\cite{MLDSA}. In our work, the theoretical security of the \textit{IPK-pq} key generation protocol mainly includes four aspects: First, we analyze whether there is a risk of collusion attack in the private seed matrix of \textit{IPK-pq} key generation protocol; Second, we analyze the impact of the matrix size on security in the \textit{IPK-pq} key generation protocol; Third, in the \textit{IPK-pq} key generation protocol, we analyze whether there is a risk of substitution attack for the declared public key; Finally, we demonstrate that the composite private key is collision resistant. Due to space limitations, please refer to the supplementary appendix for detailed security analysis.

\section{Evaluation and Analysis of IPK-pq}\label{section5}
This section primarily focuses on the performance comparison testing based on NLnetLabs's Krill\cite{Krill} (a CA and Publication server with Repo.) and Routinator\cite{Routinator} (a Relying party software), and the fundamental performance analysis of \textit{IPK-pq} running on real HSM. Utilizing the NLnetLabs's RPKI library, the insurance and verification testings for RCs and ROAs can be conducted. Furthermore, the built-in algorithm implementation based on OpenSSL $3.5.0$ within the component library can be modified to facilitate a performance comparison between \textit{IPK-pq} and the standard scheme.

To evaluate the performance of \textit{IPK-pq}  within the same RPKI framework, this work experimentally constructed a testing environment utilizing four VMware® Workstation v$15$ Pro instances on AMD Ryzen $7\ 1700$X machines, featuring $3.4$ GHz, $8$ physical and $16$ logical cores, along with $8$GB of RAM. This setup simulates the ROA insurance and verification processes in an online RPKI context. Specifically, three instances represent the APNIC, CNNIC, and ISP CAs with their respective Repo., while the fourth instance functions as the RP. This environment is designed as an independent service system, offering two primary functionalities: The first one is to input the binding information between user-defined IP segments and AS numbers, and output the corresponding RCs and signature objects such as ROAs in Krill (CA and Repo.); The second one is to input the signature objects and output the verification result of the signature object in Routinator (RP). Additionally, we conducted the functional and stress load testings within a fundamental three-layer CA structure, as shown in Figure \ref{figure 5}, allowing for adjustable CA quantities as required. In addition to the basic issuance and verification procedure, this test environment also integrates a multi-layer certificate chain that is required for authentication from the self-signed root certificate to EE certificate and individual ROA, facilitating comparisons while enabling algorithm switching between standard scheme and \textit{IPK-pq} in the entire business processing logic for comprehensive evaluation.

\subsection{Resource Certificate and ROA Generation}
In the design of the first functionality for the RCs and ROA signature, the system initializes the root RC by generating a pair of public and private keys, specifically employing an ML-DSA-$44/65/87$ key pair, respectively, and the cryptographic hash functions ($G$ and $H$ in ML-DSA) are instantiated using SHAKE$128$ and SHAKE$256$ \cite{SHA3}, respectively. APNIC and CAs jointly run \textit{``System setup stage"} and \textit{``Registration stage"} to initialize HSM and Registration record, and obtain their accompanying public key $R_{\mathsf{CA}}$, respectively. For the root RC, its IP segment $(0.0.0.0/0)$ and AS number range $([0, 2^{32}-1])$ need to be set as its authorized management domain.

For the standard scheme, the system signs subordinate CA's RC using the private key of its parent CA, computes the RC's hash value using the SHAKE$256$ algorithm, and updates the Manifest with this hash value. For the ROA object, the system generates a public/private key pair, creates an end-entity (EE) certificate, signs it using an ISP RC certificate, and employs this EE certificate to sign the CMS-formatted ROA subsequently created. For \textit{IPK-pq}, the RC is signed according to the process depicted in \textit{``Resource allocation and ROA issuance stage"}, and its hash value is similarly utilized to update the Manifest. For the ROA object, the system signs the CMS-formatted ROA created according to the process depicted in \textit{``Resource allocation and ROA issuance stage"}.
\begin{figure}[h]
	\centering
	\includegraphics[width=\linewidth]{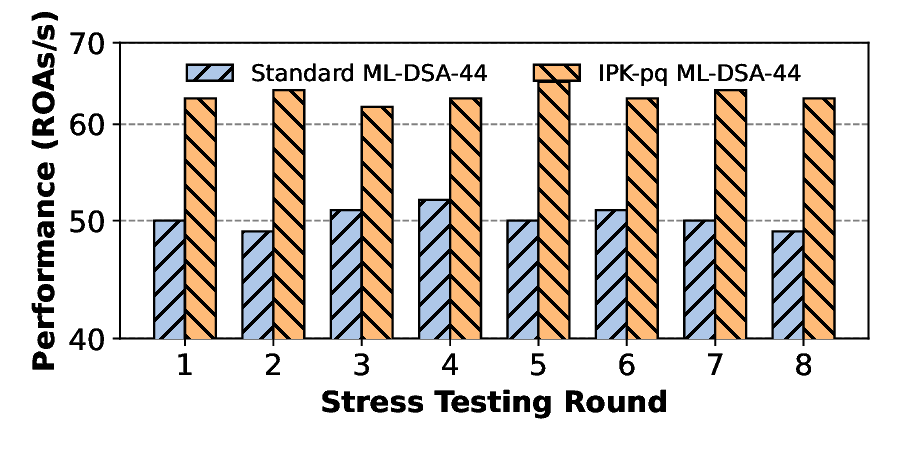}
	\caption{ROA generation performance comparison for ML-DSA-44}
	\label{figure ML-DSA-44}
\end{figure}
\begin{figure}[h]
	\centering
	\includegraphics[width=\linewidth]{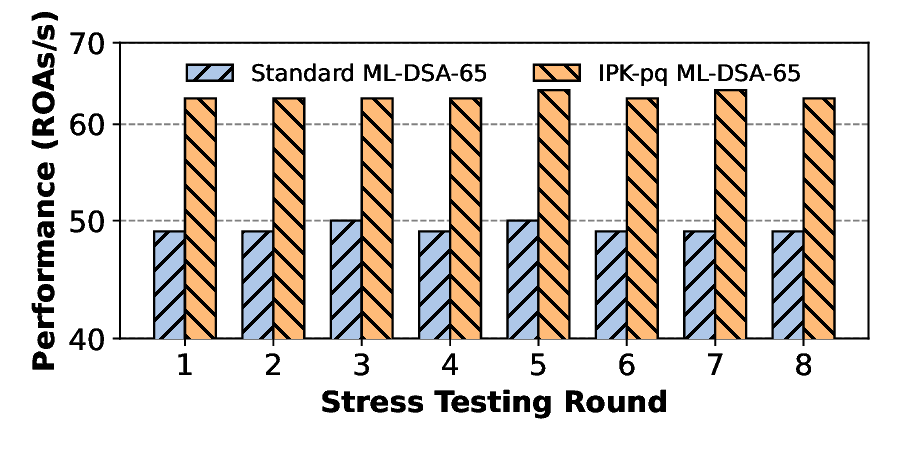}
	\caption{ROA generation performance comparison for ML-DSA-65}
	\label{figure ML-DSA-65}
\end{figure}
\begin{figure}[h]
	\centering
	\includegraphics[width=\linewidth]{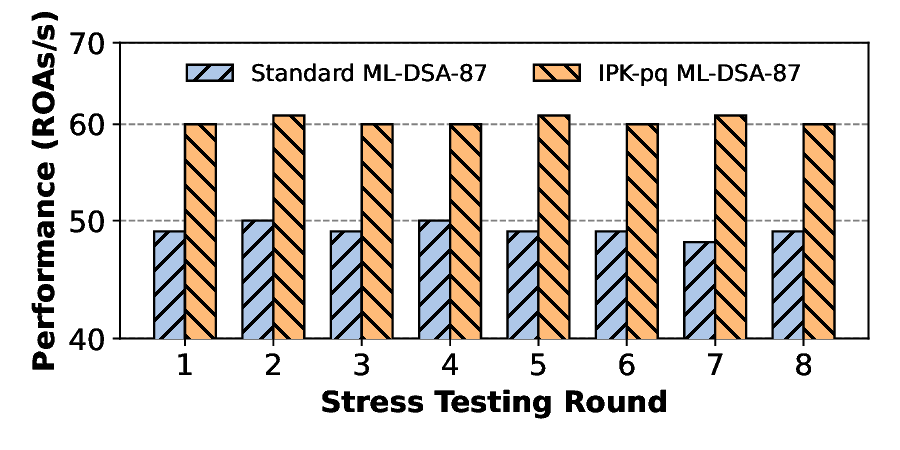}
	\caption{ROA generation performance comparison for ML-DSA-87}
	\label{figure ML-DSA-87}
\end{figure}

During the evaluation process, the system creates a new EE certificate for each different input of ROA information, and the CMS-encoded ROA object is returned to complete the entire signing process. Regarding the performance comparison between the standard scheme and \textit{IPK-pq} in RC and ROA generation, this work only focuses on \textit{"the ROA issuance efficiency reflected in the whole process from signing the root RC to issuing the ROA object"}. For ML-DSA-$44$, ML-DSA-$65$, and ML-DSA-$87$,  continuous stress testing conducted over $8$ rounds, was performed to compare the number of ROA object signatures that can be generated per second. Under the same testing conditions, the results are shown in Figure \ref{figure ML-DSA-44}, \ref{figure ML-DSA-65} and \ref{figure ML-DSA-87}, where the \textit{IPK-pq} can generate approximately $63$, $62$ and $60$ ROAs per second, respectively, while the standard scheme generates around $50$, $50$ and $49$ ROAs per second, respectively. \textit{IPK-pq} outperforms the standard scheme in the overall signature process,  with a performance improvement of approximately $1.2$ times.
Performance analysis indicates that the low efficiency of the standard scheme in generating RCs and ROAs is primarily due to the number of signatures required in the ROA generation process. This process involves two complete signing operations: ($1$) the ML-DSA signature on the CMS message digest after ROA encoding, ensuring the integrity of the ROA, and ($2$) the signature on the EE ML-DSA public key by the parent CA, verifying the validity of the ROA object. In the standard scheme, for ML-DSA-$44$, ML-DSA-$65$, and ML-DSA-$87$, these two signatures account for an average of $38.18\%$ ($19.22\%$ and $18.96\%$, respectively), $38.70\%$ ($19.37\%$ and $19.33\%$, respectively), $38.97\%$ ($19.50\%$ and $19.47\%$, respectively) of the total testing time, respectively. In contrast, for \textit{IPK-pq}, there is only the first signature for ROA, with no second signature for the EE certificate. For ML-DSA-$44$, ML-DSA-$65$, and ML-DSA-$87$,  the first signature only accounts for an average of $24.83\%$, $24.93\%$, and $25.11\%$ of the total time, respectively.

\subsection{Resource Certificate and ROA Verification}
In the design for the second functionality concerning RC and ROA verification, we start from obtaining ROA, which is verified systematically layer by layer according to the trusted RC certificate chain.
For \textit{IPK-pq}, the system directly decodes ROAs, and extracts the signature information such as the signer CA's identity $ID_{\mathsf{CA}}$, accompanying public key $R_{\mathsf{CA}}$, and signature $\sigma$ of ROAs, verifies whether the $ID_{\mathsf{CA}}$ and public key $R_{\mathsf{CA}}$ are valid in Registration record table, then verifies $\sigma$ to check if it can pass the signature verification according to the process depicted in \textit{``ROA Verification stage"}. For the standard scheme, the system repeats the verification process until the root RC is reached, if RC is a root one, which means it does not contain any valid parent RC. Then, the system uses this RC directly to compare the algorithm with the one in the trust RC included in the built-in TAL\cite{PKICRFC, IETFRFC6916} to check if it is valid. If the root RC verification is successful, the entire verification process is successful; otherwise, a failure message is returned.

\begin{figure}[h]
	\centering
	\includegraphics[width=\linewidth]{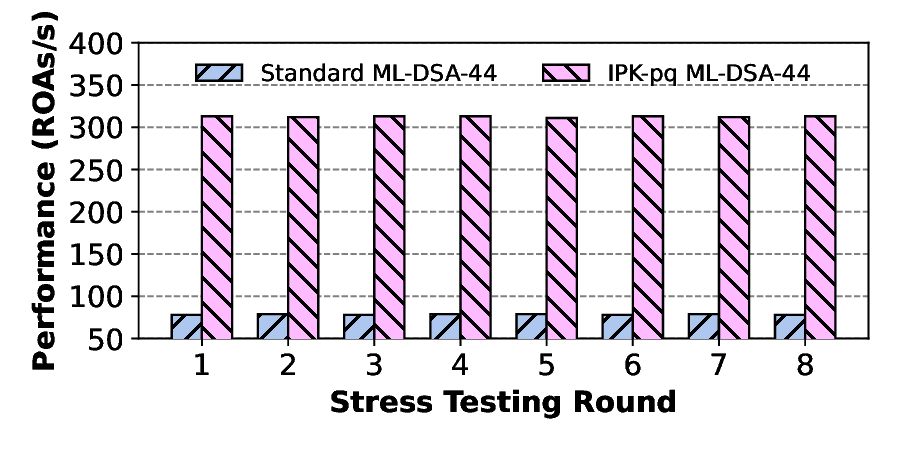}
	\caption{ROA verification performance comparison for ML-DSA-44}
	\label{figure ML-DSA-44-VERI}
\end{figure}
\begin{figure}[h]
	\centering
	\includegraphics[width=\linewidth]{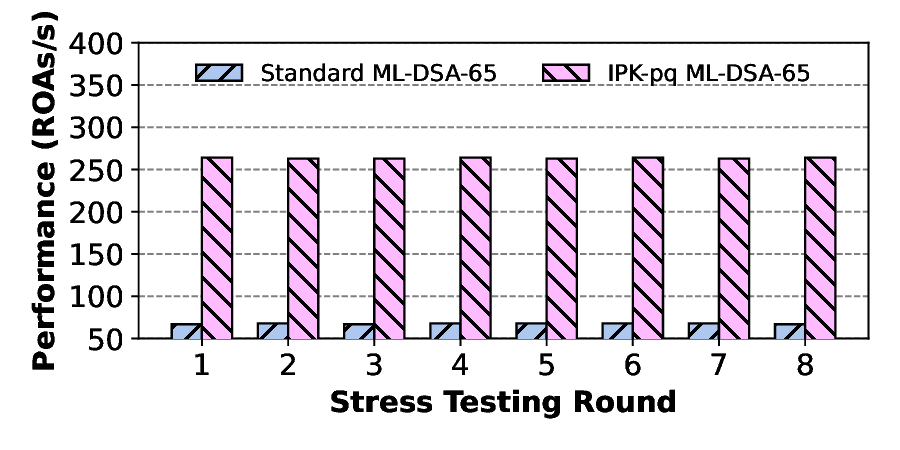}
	\caption{ROA verification performance comparison for ML-DSA-65}
	\label{figure ML-DSA-65-VERI}
\end{figure}
\begin{figure}[h]
	\centering
	\includegraphics[width=\linewidth]{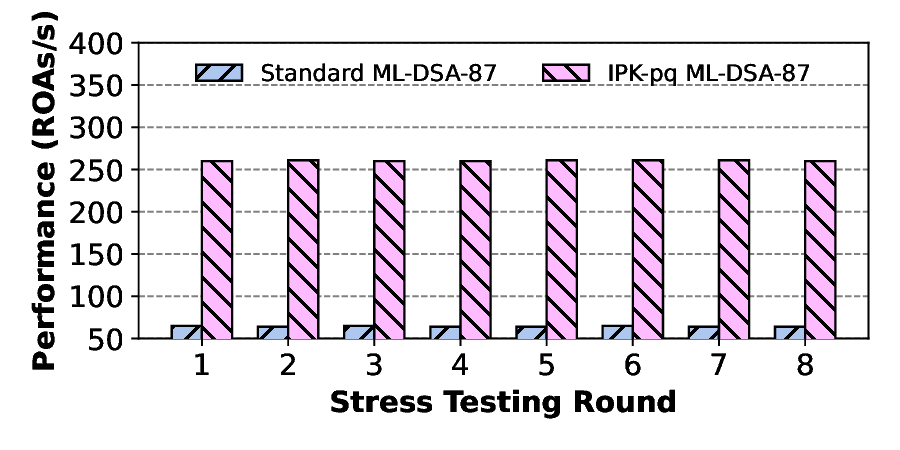}
	\caption{ROA verification performance comparison for ML-DSA-87}
	\label{figure ML-DSA-87-VERI}
\end{figure}

\begin{table}[!tbp]
	\caption{The basic size comparison between IPK-pq and the standard scheme}
	\label{IPK-pq and standard scheme comparison}
	\centering
	\begin{tabular}{ccc}
		\toprule
		RPKI Algorithm&Private Key Length&Public Key Length\\
		\midrule
		 IPK-pq ML-DSA-$44$& $2,560 $B& $(32+Len(ID_{\mathsf{CA}}))$B   \\
		 IPK-pq ML-DSA-$65$& $4,032 $B& $(32+Len(ID_{\mathsf{CA}}))$B   \\
		 IPK-pq ML-DSA-$87$& $4,896 $B& $(32+Len(ID_{\mathsf{CA}}))$B   \\
		Standard ML-DSA-$44$ & $2,560$B &  $1,312$B  \\
		Standard ML-DSA-$65$ & $4,032$B &  $1,952$B  \\
		Standard ML-DSA-$87$ & $4,896$B &  $2,592$B  \\
		\bottomrule
	\end{tabular}
\end{table}

\begin{figure}[h]
	\centering
	\includegraphics[width=\linewidth]{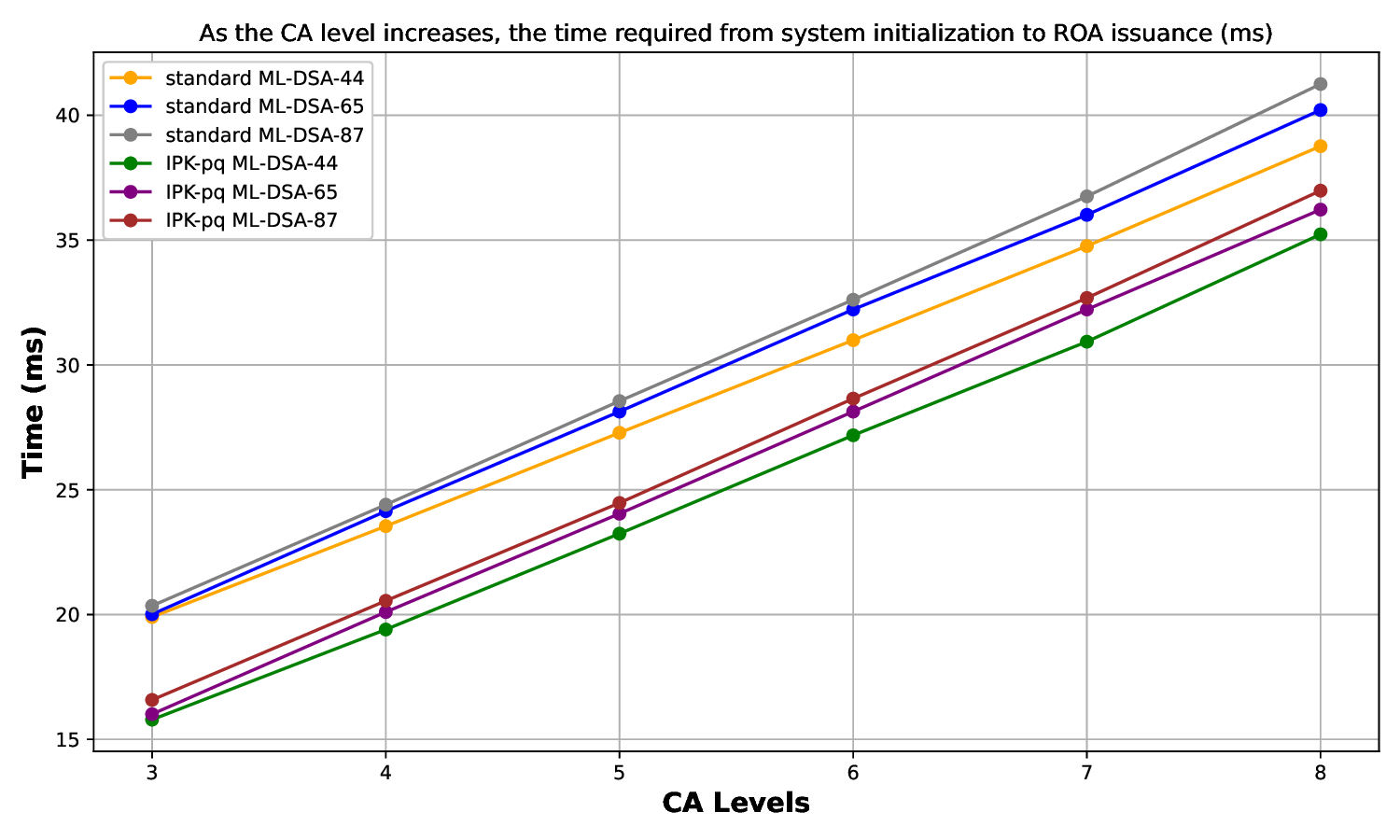}
	\caption{As the CA level increases from $3$ to $8$ layers, the time required for ROA generation after system initialization (ms)
	}
	\label{figure INC-ROA-GEN}
\end{figure}

\begin{figure}[h]
	\centering
	\includegraphics[width=\linewidth]{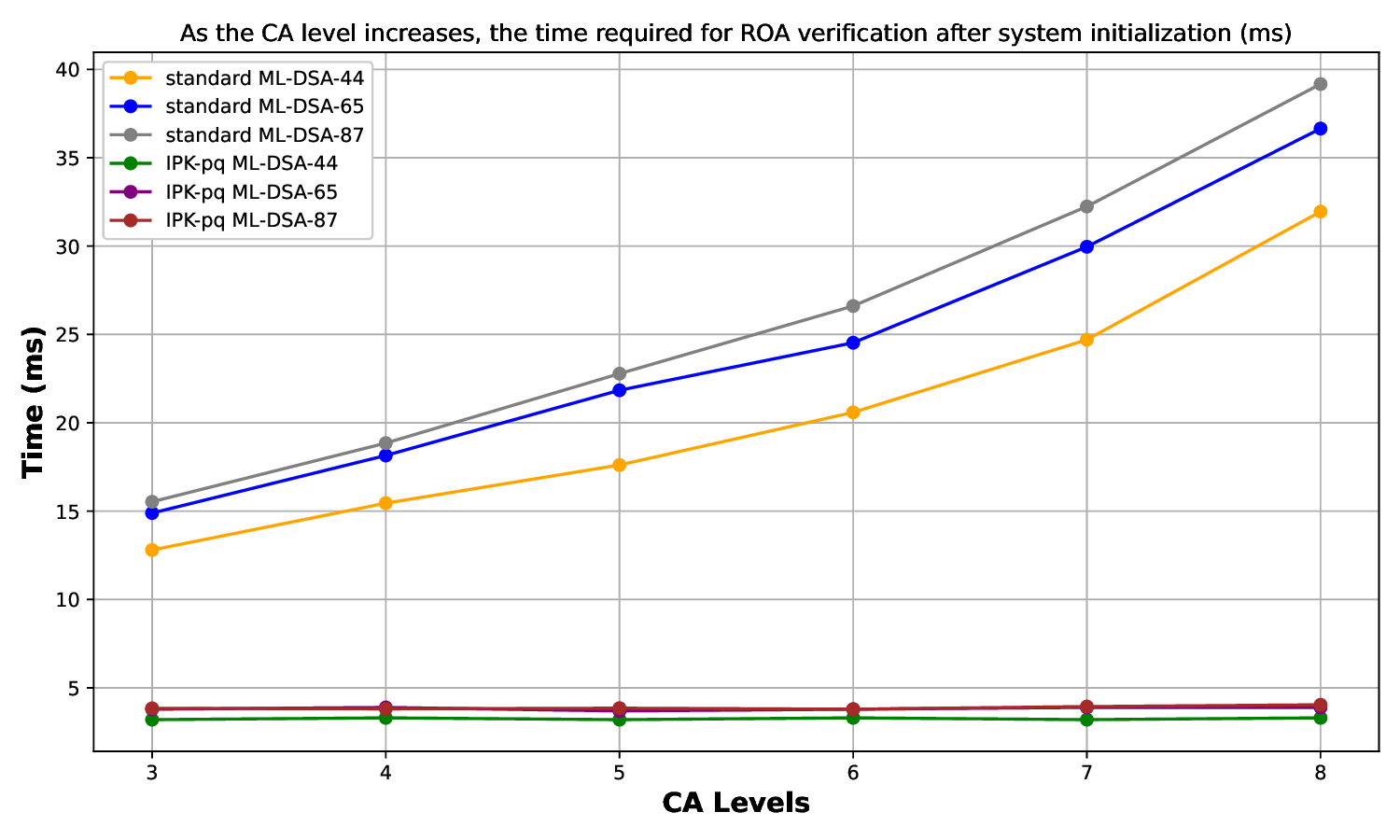}
	\caption{As the CA level increases from $3$ to $8$ layers, the time required for ROA verification after system initialization (ms)
	}
	\label{figure INC-ROA-VER}
\end{figure}

For ML-DSA-$44$, ML-DSA-$65$, and ML-DSA-$87$, Figure \ref{figure ML-DSA-44-VERI}, \ref{figure ML-DSA-65-VERI} and \ref{figure ML-DSA-87-VERI} show the performance comparison of \textit{"ROA validation"}  for a three-layer CA structure, where \textit{IPK-pq} can verify approximately $312$, $263$ and $260$ ROAs per second, respectively, while the standard scheme verifies around $78$, $67$ and $64$ ROAs per second, respectively. Performance analysis from stress testing (8 rounds) demonstrates that \textit{IPK-pq} achieves verification efficiency approximately 4 times higher than the standard scheme for ROAs and RCs. This improvement is largely attributed to the elimination of the need for RPs to download the full RC certificate chain (containing all RCs) from APNIC to the CA issuing the ROA and EE certificate, significantly enhancing verification performance. While the standard scheme requires at least two signature operations per ROA, \textit{IPK-pq} completes the process with only one. Additionally, as the RC certificate chain grows in the standard scheme, compared with ROA signature generation, the cost of signature verification increases sharply, as shown in Figure \ref{figure INC-ROA-GEN} and \ref{figure INC-ROA-VER}. It is clear that during ROA generation, both approaches have a complexity of \(\mathcal{O}(n)\). However, in ROA verification, the standard scheme has a complexity of \(\mathcal{O}(n)\), while \textit{IPK-pq} has a complexity of \(\mathcal{O}(1)\) for ML-DSA-$44$, ML-DSA-$65$, and ML-DSA-$87$. The comparison also includes operations such as RC downloading, ASN.1 data decoding, and data extraction from RCs and ROAs.

\begin{table*}
	\begin{center}
		\caption{The performance comparison between IPK-pq and standard scheme in a real-world deployment instance}
		\label{remoteop}
		\begin{threeparttable}
			
			\begin{tabular}{p{3cm}|p{3cm}|p{3cm}|p{3cm}|p{3cm}}
				\toprule
				Role & APNIC & CNNIC& ISP& RP\\
				$(Instance's\ Location)$ & $(Sydney, Australia)$ & $(Beijing, China)$& $(Beijing, China)$& $(Ningxia, China)$ \\
				\hline
				Latency (ms)  & $\sim$$238$ (Sydney$\Leftrightarrow$Beijing) & $\sim$1 (Beijing $\Leftrightarrow$ Beijing)&\multicolumn{2}{l}{$\sim$$19.5$ (Beijing $\Leftrightarrow$ Ningxia)}\\
				\hline
				Bandwidth (Mb/s) & $100$ (Sydney $\Leftrightarrow$ Beijing) &$100$ (Beijing $\Leftrightarrow$ Beijing)&\multicolumn{2}{l}{$100$ (Beijing $\Leftrightarrow$ Ningxia)}\\
				\hline
				\multirow{3}{*}{ROA Avg. Generation } & \multicolumn{2}{l|}{IPK-pq ML-DSA-44: $60$ $\pm$ $12$ ROAs per second.}  &\multicolumn{2}{l}{standard ML-DSA-44: $45$ $\pm$ $11$ ROAs per second.\tnote{1}} \\
				& \multicolumn{2}{l|}{IPK-pq ML-DSA-65: $57$ $\pm$ $10$ ROAs per second.} &\multicolumn{2}{l}{standard ML-DSA-65: $44$  $\pm$ $13$ ROAs per second.\tnote{1}}\\
				&\multicolumn{2}{l|}{IPK-pq ML-DSA-87: $55$ $\pm$ $11$ ROAs per second.} &\multicolumn{2}{l}{standard ML-DSA-87: $44$ $\pm$ $10$ ROAs per second.\tnote{1}} \\
				\hline
				\multirow{3}{*}{ROA Avg. Verification } & \multicolumn{2}{l|}{IPK-pq ML-DSA-44: $262$ $\pm$ $5$ ROAs per second.}&\multicolumn{2}{l}{standard ML-DSA-44: $25$ $\pm$ $10$ ROAs per second.\tnote{1}} \\
				& \multicolumn{2}{l|}{IPK-pq ML-DSA-65: $223$ $\pm$ $8$ ROAs per second.}&\multicolumn{2}{l}{standard ML-DSA-65: $24$ $\pm$ $13$ ROAs per second.\tnote{1}} \\
				&\multicolumn{2}{l|}{IPK-pq ML-DSA-87: $215$ $\pm$ $8$ ROAs per second.}&\multicolumn{2}{l}{standard ML-DSA-87: $22$ $\pm$ $12$ ROAs per second.\tnote{1}} \\
				\bottomrule
			\end{tabular}
			\begin{tablenotes}
				\footnotesize
				\item[1] APNIC's RC is usually cached as RIR root RC, performance data changes depends on whether the intermediate RCs are cached, as well as the number of CA layers contained in a specific deployment instance and its network latency and bandwidth.
			\end{tablenotes}
		\end{threeparttable} 		
	\end{center}
\end{table*}

\begin{figure}[h]
	\centering
	\includegraphics[width=0.6\linewidth]{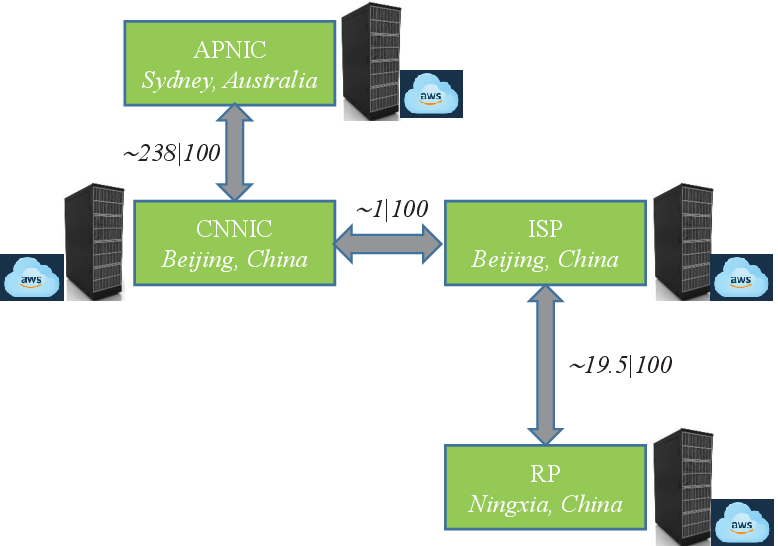}
	\caption{$Latency|Bandwidth$ between regions, where latency is in milliseconds and bandwidth is in Mbits/s. Since AWS has no cloud Region in Brisbane, and APNIC is actually located in Brisbane, this evaluation uses a node instance from the Sydney Region to represent APNIC, to meet remote testing conditions as closely as possible.}
	\label{figure remoteops}
\end{figure}

The elimination of RC certificate chain downloads significantly improves verification performance, resulting in higher overall efficiency compared to the standard scheme, aligning with theoretical analysis. In RPKI operations, CA primarily focuses on generating RCs and ROAs, while RP is responsible for verifying these objects. After validation, RP passes the verified ROAs to BGP routers and other devices for loading. Once generated, RC and ROA objects typically undergo infrequent changes, shifting the emphasis to verifying a large number of routing entries. Thus, enhancing verification efficiency is crucial for the large-scale adoption of RPKI. To evaluate the ROA issuance efficiency  of remote operations throughout \textit{"the whole process from signing the root RC to issuing the ROA object"}, as well as the verification efficiency during \textit{"ROA validation"}, within a limited budget, we deployed the four-node setup on Amazon AWS c$5.2$xlarge instances equipped with Intel Xeon Cascade Lake CPUs ($3.8$ GHz, $16$ GB RAM). The configuration and performance data are shown in Table \ref{remoteop}. The configured instances, as well as the latency and bandwidth between instances, are shown in Figure \ref{figure remoteops}.

\begin{figure}[h]
	\centering
	\includegraphics[width=\linewidth]{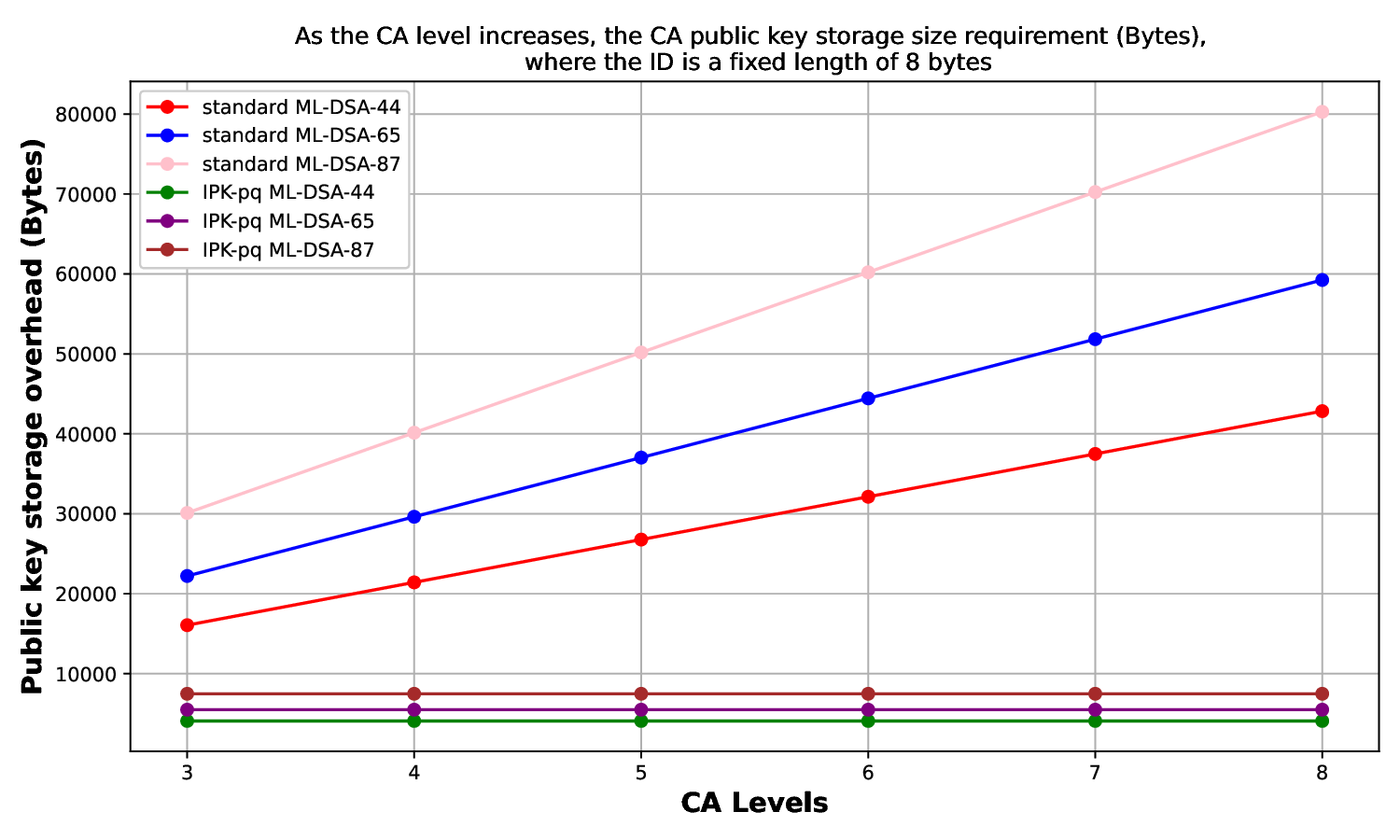}
	\caption{As the CA level increases from $3$ to $8$ layers, the CA's public key storage size requirement (Bytes), where the ID is a fixed length of 8 bytes
	}
	\label{figure Storage}
\end{figure}

\begin{figure}[h]
	\centering
	\includegraphics[width=\linewidth]{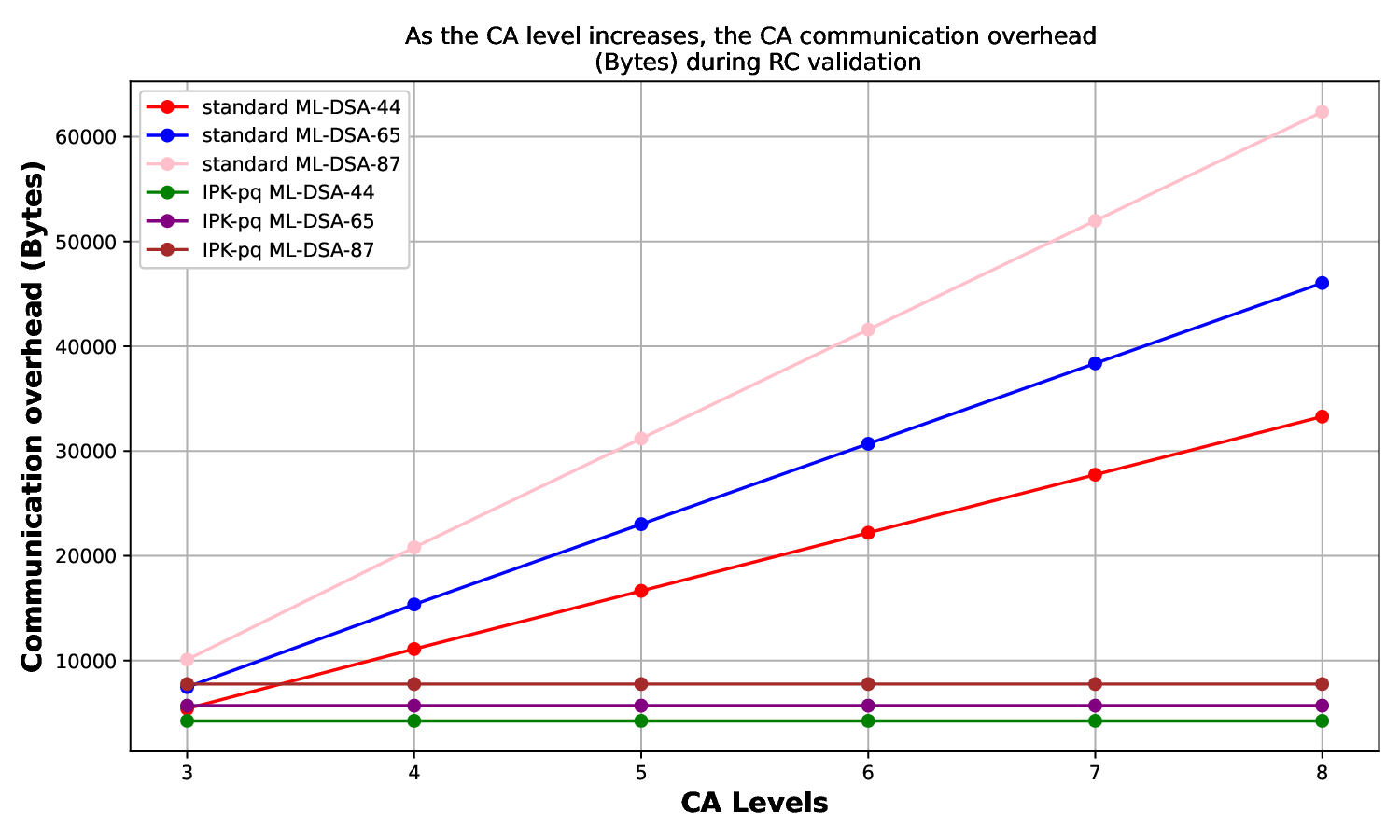}
	\caption{As the CA level increases from $3$ to $8$ layers, the CA's communication overhead (Bytes), where the ID is a fixed length of 8 bytes
	}
	\label{figure Communication}
\end{figure}

Obviously, \textit{IPK-pq} enhances key distribution efficiency and improves system performance, as shown in Table \ref{IPK-pq and standard scheme comparison} (partially cited from \cite{MLDSA}) and Table \ref{remoteop}. In our work, the authentication of a signature object requires only one cryptographic operation. This approach offers three key advantages:
($1$) It is compatible with the existing ML-DSA-based PKI while eliminating the need for additional RC certificate chains for verification. This significantly enhances verification efficiency, especially in scenarios involving frequent remote RC retrievals.
($2$) With the identity-based key management scheme, the CA's private keys are not arbitrary but require collaboration between the Key Center and the CA during initialization. The Key Center cannot access the full ML-DSA private keys of the CAs, thereby resolving the key escrow problem. In resource-constrained scenarios, the public random seed matrix $\mathbf{M}_{\rho}^{\text{pub}}$ can be replaced with a single random seed to further reduce the size of the public key file $File_{\mathsf{PK}}$.
($3$) In \textit{IPK-pq}, ML-DSA public keys are stored in a public key file managed by the Key Center, enabling download, online retrieval, or query as needed. For standard ML-DSA-$44$, ML-DSA-$65$, and ML-DSA-$87$, the certificate sizes are $5,355$B, $7,406$B, and $10,035$B, respectively; in contrast, for \textit{IPK-pq}, they are reduced to $4,083$B, $5,494$B, and $7,483$B, respectively.

Figure \ref{figure Storage} compares the storage requirements for the CA's public key (included in RCs) during RC validation between the standard scheme and \textit{IPK-pq}, assuming \textit{IPK-pq} operates with online queries. In the standard scheme, public key storage for the lowest-level CA increases significantly with the number of CA tiers. For standard ML-DSA-$44$, the storage requirement escalates from $16,065$B to $42,840$B; for standard ML-DSA-$65$, from $22,228$B to $59,248$B; and for standard ML-DSA-$87$, from $30,105$B to $80,280$B. In contrast, \textit{IPK-pq} only requires storage for the RC corresponding to the fixed identity $ID_{\mathsf{CA}}$ and its public key $R_{\mathsf{CA}}$, maintaining constant storage costs of approximately $4,083$B, $5,494$B, and $7,483$B, respectively.

Under the same conditions, Figure \ref{figure Communication} compares the communication overhead during RC validation between the two schemes, assuming the root RC is locally cached in the standard scheme. In the standard scheme, the communication overhead for the lowest-level CA increases linearly with the number of CA tiers. For standard ML-DSA-$44$, the overhead rises from $5,409$B to $33,287$B; for standard ML-DSA-$65$, from $7,460$B to $46,036$B; and for standard ML-DSA-$87$, from $10,089$B to $62,378$B. Conversely, \textit{IPK-pq} requires querying only the ML-DSA RC containing the public key corresponding to $ID_{\mathsf{CA}}$, resulting in stable communication overheads of approximately $4,230$B, $5,692$B, and $7,753$B, respectively. Overall, both the public key storage overhead and communication complexity for ROA verification in the standard scheme exhibit a complexity of \(\mathcal{O}(n)\), whereas \textit{IPK-pq} achieves a complexity of \(\mathcal{O}(1)\). This highlights the significant advantages of \textit{IPK-pq} in terms of scalability and efficiency.

Another critical advantage of \textit{IPK-pq} lies in its superior amortized communication performance. In traditional RPKI, the certificate chain must be retrieved and validated for every routing object, leading to a linear increase in communication overhead proportional to the CA depth. In contrast, \textit{IPK-pq} shifts the primary data transmission cost to the system initialization phase. The public parameter file $File_{PK}$ is downloaded once by the Relying Party and cached locally, or only requires an online query. Subsequent verifications only require minimal incremental updates or specific entry queries, making the marginal communication cost for each verification event negligible. As the number of verification requests ($N$) increases, the average communication overhead per verification approaches zero (i.e., $\mathcal{O}(1/N)$), or only \(\mathcal{O}(1)\) online query communication overhead, demonstrating high scalability for hyperscale IIoT and global routing scenarios.

To verify the practicality of RPKI based on \textit{IPK-pq}, we evaluate whether the scheme meets the efficiency requirements of real-world RIR operations. We accessed the publicly available historical RPKI data maintained by RIPE NCC\footnote{https://ftp.ripe.net/rpki/. This database contains the daily historical data for all regions that are publicly accessible.}, which includes daily archives of all five RIR repositories since $2011$. We utilized historical data from March $11$, $2015$, to Jan $19$, $2026$, consistent with the methodology in \cite{KrisShrishak}, where Shrishak et al. analyzed the daily fluctuation of ROAs. Based on estimates derived from this data (updated from $2020$ to $2026$), the real-world RPKI production system requires approximately $8$K signatures per day on average, with occasional peaks due to ROA re-issuance. The proposed \textit{IPK-pq} scheme is capable of generating over $250$ signatures per second, equating to approximately $21,600$K signatures per day. This throughput is roughly $2,700$ times higher than the current average daily demand. These results demonstrate that \textit{IPK-pq} provides ample performance margins to accommodate the increasing adoption rate of RPKI, ensuring robust scalability.

\subsection{IPK-pq based on Hardware HSM and Its Evaluation}
Furthermore, to validate the system's performance in a hardware-accelerated environment, building on the prior stress testing assessment, we implemented \textit{IPK-pq} utilizing an HSM equipped with an NXP C$293$ PCIe Card\footnote{The C29x PCI Express (PCIe) adapter platform serves as an optimal hardware and software development environment tailored for high-performance networking applications, such as Unified Threat Management, Application Delivery Controllers, and Hardware Security Modules. For more information, please refer to https://www.nxp.com.cn/products/C29x.}. The C$293$ integrates three high-performance cryptographic coprocessors optimized for public key operations, providing hardware acceleration for protocols like IKE, SSL, DNSSEC, and secure BGP, while supporting the longer key lengths required by modern cryptography. Additionally, the C$293$ coprocessor supports open micro-instructions for mathematical operations such as big number modular arithmetic, enabling the customized implementation of matrix-vector multiplication and other polynomial ring operations required by ML-DSA. The configuration parameters on HSM and \textit{IPK-pq} are outlined in the Table \ref{HSM} below.
It is important to note that the C293 coprocessor does not natively support SHAKE algorithms. Consequently, for this prototype implementation, SHAKE$128$ and SHAKE$256$ in the ML-DSA algorithm were substituted with SHA$256$ to leverage hardware acceleration\footnote{In the construction of \textit{IPK-pq} based on hardware HSM, substituting SHAKE (XOF) with SHA256 (hash) serves solely to validate the feasibility of an independent HSM hardware acceleration prototype. During actual deployment, users should utilize hardware that supports SHAKE.}. While this modification facilitates performance evaluation, future production deployments would utilize hardware fully compliant with FIPS 204. In the \textit{IPK-pq} workflow, the HSM utilizes a PRNG to generate the public and private seed matrices during initialization. During the registration phase, it derives private seed components to generate private keys for the CA. Finally, in the RC resource allocation and ROA issuance processes, the HSM accelerates the ML-DSA algorithms for signature generation and verification.

\begin{table}[h]
	\centering
	\caption{C$293$ HSM System Configuration}
	\label{HSM}
	\begin{tabular}{@{}ll@{}}
		\toprule
		\textbf{System Parameter}       & \textbf{Value}                            \\ \midrule
		CPU                     & $1.2$GHZ, 32b, e500v2 core \\
		I and D caches                     & 32KB, respectively \\ 			
		SEC                     & $400$MHZ      \\
		DDR$3$                     & $512$MB with ECC, $1.2$GHZ                           \\
		SRAM                     & $512$KB                                    \\
		L$2$ Cache                 & $512$KB                                    \\
		NAND Flash               & $4$GB                                      \\
		Operating Mode                     & PK Calculator                                \\
		PCIe Gen $2.0$            & x$4$                                       \\
		Matrix Parameter \( m \) & $32$                                       \\
		Matrix Parameter \( h \) & $32$                                      \\
		$ID_{\_CA}$ length        & $8$B                                       \\
		Other ML-DSA Parameters         & Refer to \cite{MLDSA}                    \\ \bottomrule
	\end{tabular}
\end{table}

\begin{figure}[h]
	\centering
	\includegraphics[width=\linewidth]{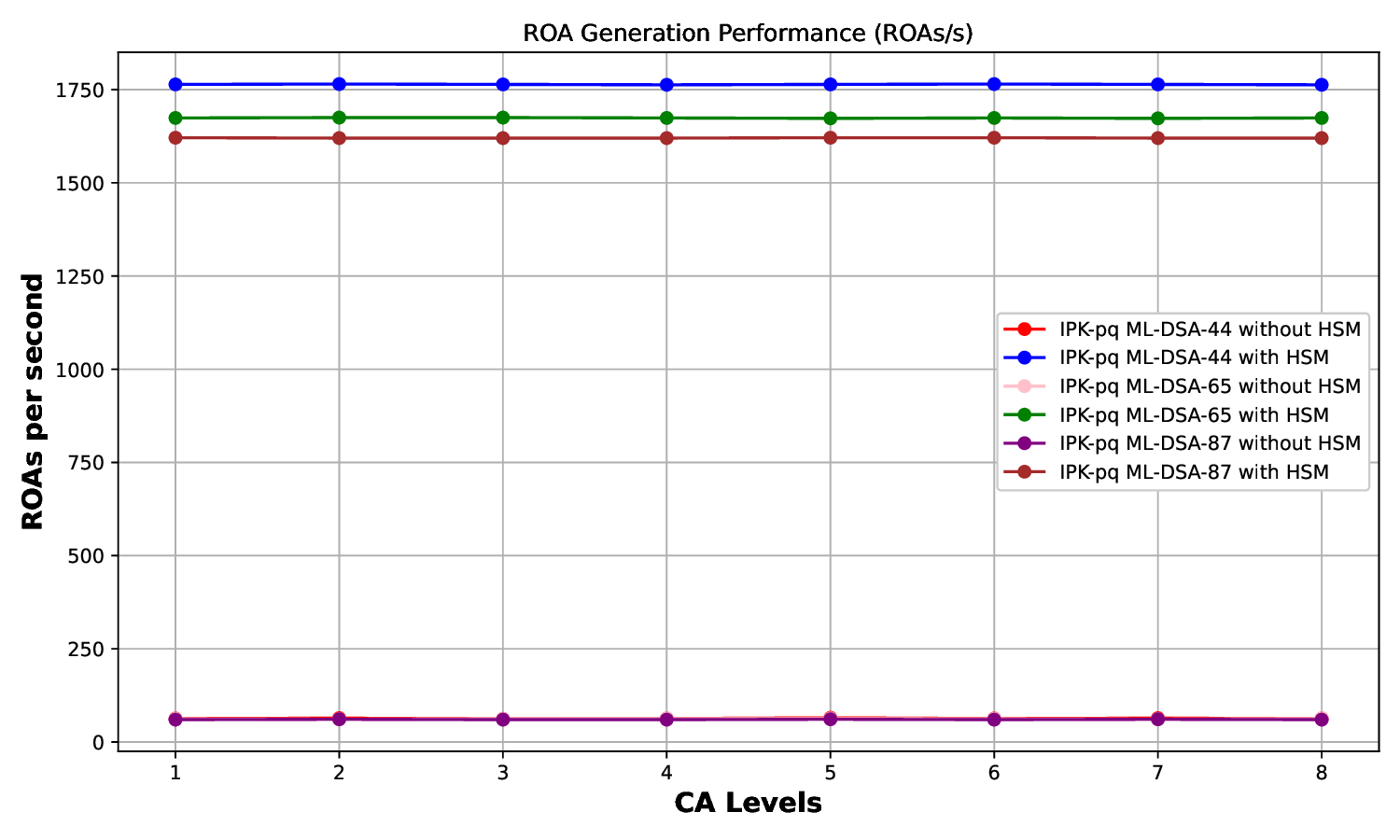}
	\caption{IPK-pq ROA generation performance with and without HSM}
	\label{figure HSMSIG}
\end{figure}
\begin{figure}[h]
	\centering
	\includegraphics[width=\linewidth]{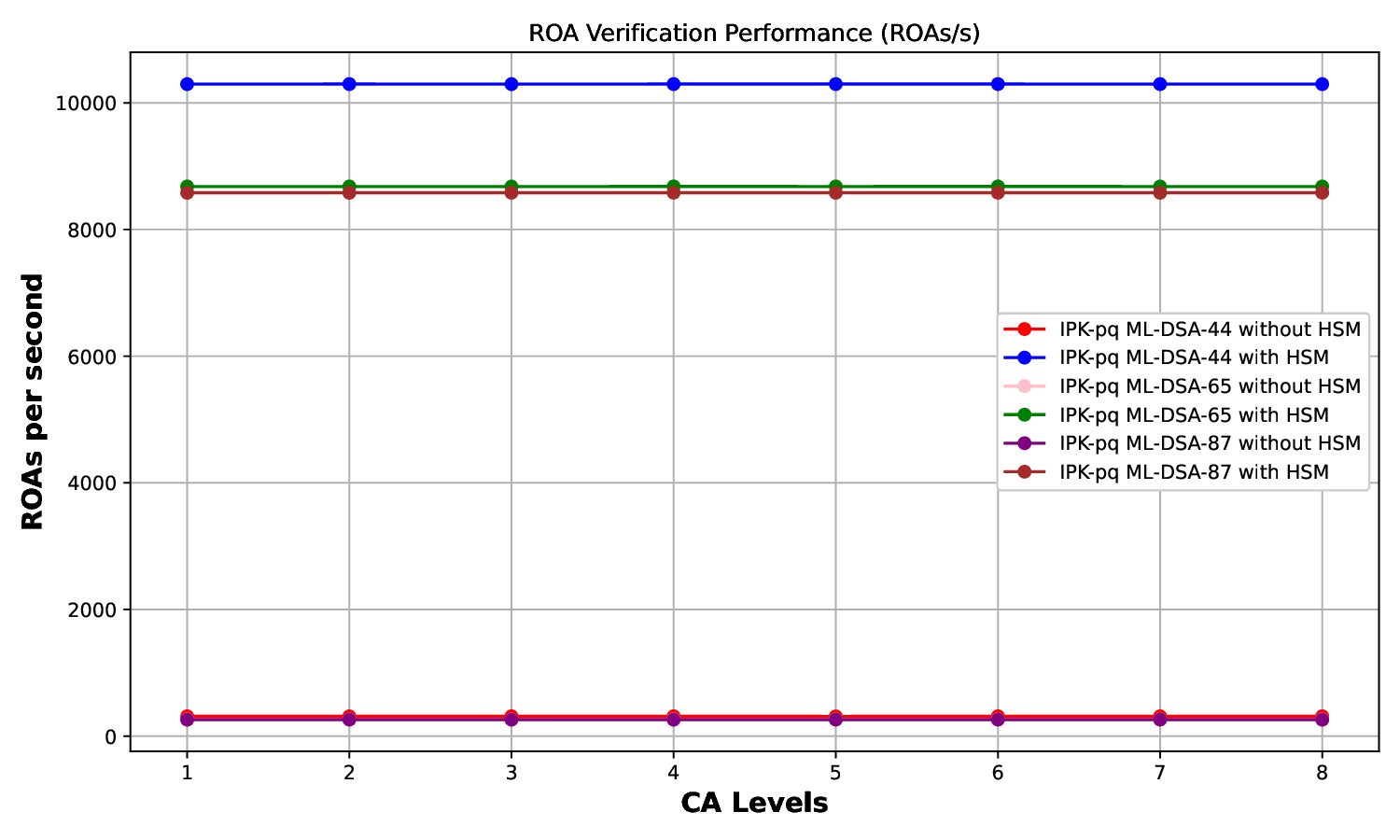}
	\caption{IPK-pq ROA verification performance with and without HSM}
	\label{figure HSMVER}
\end{figure}

Building on the previous comparative testing, the issuance and verification processes of RCs and ROAs were further modified to implement the ML-DSA ciphers using an HSM, with data processed and verified layer by layer according to the hierarchical structure. Figures \ref{figure HSMSIG} and \ref{figure HSMVER} compare the performance of \textit{IPK-pq} for ROA generation and verification with and without HSM over eight rounds, respectively. For a basic three-layer CA structure as shown in Figure \ref{figure 5}, we focus more on the performance and efficiency of ROA issuance and verification, encompassing the entire process from signing the root RC to issuing the ROA object, rather than the time for a single signature. It is clear that, ROA generation performance improved from $63$ TPS  (transactions per second, that is, ROAs/s), $62$ TPS, and $60$ TPS to $1,765$ TPS, $1,674$ TPS, and $1,621$ TPS (a $\sim$$27$x increase), while verification performance rose from $312$ TPS, $263$ TPS, and $260$ TPS to $10,296$ TPS, $8,679$ TPS, and $8,580$ TPS (a $\sim$$33$x increase), respectively. The results demonstrate that HSM can significantly enhance the performance and security in handling large-scale RCs and ROAs within RPKI. Further optimizations, such as improving hardware drivers and FIFO scheduling algorithms, could achieve even greater performance gains.


\section{Conclusion}\label{section6}
To meet the need for efficient cryptographic systems in post-quantum PKI, we introduce \textit{IPK-pq}, the first identity and HSM-based, certificate chain-free post-quantum PKI scheme, which eliminates the need for users to store large numbers of certificates or download certificate chains, significantly reducing the overheads of certificate validation, storage and transmission for users. Designed for high efficiency in large-scale deployments, it is also well-suited for IIoT scenarios containing devices with limited storage. Using RPKI as an example, extensive testing demonstrates that \textit{IPK-pq} significantly improves RP validation efficiency. Compatible with various ML-DSA-based PKI systems, it offers flexibility and adaptability for diverse deployment scenarios. Moreover, \textit{IPK-pq} enables RPKI to scale efficiently, lowering infrastructure and operational costs. With its robust, efficient, and scalable design, \textit{IPK-pq} modernizes RPKI, making it better suited for secure, high-performance, and large-scale internet routing.

\section*{Acknowledgment}
The authors wish to thank the editors and anonymous reviewers for their valuable comments that contributed to the improved quality of this paper. This work is supported in part by the Major Key Project of PCL under Grant PCL2025A13.



\ifCLASSOPTIONcaptionsoff
  \newpage
\fi

\vspace{-20pt}
\begin{IEEEbiographynophoto}{Penghui Liu}
	 worked as a Senior Security Expert and Technical Manager of Chipset Security Technology Department at SMiT, has over 22 years of experience in research and application of cryptographic technologies. Now he works in the Department of New Networks, Pengcheng laboratory, Shenzhen, China. He is also a member of ACM.
\end{IEEEbiographynophoto}
\vspace{-20pt}
\begin{IEEEbiographynophoto}{Yi Niu}
	works as a Technical Expert and Founder at Nanjing Xunshi Data Technology Co., Ltd., has over 30 years of experience in the application of lightweight cryptographic technology, improving the IPK system, and focusing on the research of post quantum cryptographic technology. He is also a member of ACM.
\end{IEEEbiographynophoto}
\vspace{-20pt}
\begin{IEEEbiographynophoto}{Xiaoxiong Zhong}
	obtained his Ph.D. degree of Engineering in computer science and technology, Harbin Institute of Technology in 2015. Now he works in the Department of New Networks, Pengcheng laboratory, Shenzhen, China. He is also a member of ACM.
\end{IEEEbiographynophoto}
\vspace{-20pt}
\begin{IEEEbiographynophoto}{Jiahui Wu}
	obtained her Ph.D. degree in signal and information processing from Southwest University, Chongqing, China, in 2021.
    Now she works in the Department of New Networks, Pengcheng laboratory, Shenzhen, China.
    Her research interests include cryptography, privacy-preserving machine learning, and cloud computing security.
\end{IEEEbiographynophoto}
\vspace{-20pt}
\begin{IEEEbiographynophoto}{Weizhe Zhang}
	(Senior Member, IEEE) obtained his Ph.D. degree of Engineering in computer science and technology, Harbin Institute of Technology in 2006. Now he works in the Department of New Networks, Pengcheng laboratory, Shenzhen, China. He is also Senior member of ACM.
\end{IEEEbiographynophoto}
\vspace{-20pt}
\begin{IEEEbiographynophoto}{Kaiping Xue}
	(Senior Member, IEEE) obtained the bachelor's degree from the Department of Information Security, University of Science and Technology of China (USTC), in 2003, and the Ph.D. degree from the Department of Electronic Engineering and Information Science (EEIS), USTC, in 2007. From May 2012 to May 2013, he was a Postdoctoral Researcher with the Department of Electrical and Computer Engineering, University of Florida. Currently, he is a Professor with the School of Cyber Science and Technology, USTC. He is also the Director of the Network and Information Center, USTC. His research interests include next-generation Internet architecture design, transmission optimization, and network security. He is an IET Fellow. His work won best paper awards in IEEE MSN 2017 and IEEE HotICN 2019, IEEE ICC 2025, the Best Paper Honorable Mention in ACM CCS 2022, the Best Paper Runner-Up Award in IEEE MASS 2018, and the best track paper in MSN 2020. He serves on the editorial board for several journals, including IEEE TANSACTIONS ON INFORMATION FORENSICS AND SECURITY, IEEE TRANSACTIONS ON DEPENDABLE AND SECURE COMPUTING, IEEE TRANSACTIONS ON WIRELESS COMMUNICATIONS, and IEEE TRANSACTIONS ON NETWORK AND SERVICE MANAGEMENT. He has served as a guest editor for many reputed journals/magazines. He has also served as the track/symposium chair for some reputed conferences.
\end{IEEEbiographynophoto}
\vspace{-20pt}
\begin{IEEEbiographynophoto}{Bin Xiao}
	(Fellow, IEEE) obtained the BS and MS degrees in electronics engineering from Fudan University, China, and the PhD degree in computer science from the University of Texas at Dallas, Richardson, X. He is currently an associate professor with the Department of Computing, The Hong Kong Polytechnic University. He has more than ten years research experience with cyber security, and currently focuses on blockchain technology and AI security. He has published more than 100 technical papers in international top journals and conferences. Currently, he is an associate editor of the Journal of Parallel and Distributed Computing (JPDC) and the vice chair of IEEE ComSoc CISTC committee. He has been the symposium co-chair of IEEE ICC2020, ICC 2018 and Globecom 2017, and the general chair of IEEE SECON 2018.
\end{IEEEbiographynophoto}
\vfill

\end{document}